\documentclass[aps,pra,preprint,groupedaddress]{revtex4-2}

\usepackage[dvips]{graphicx}
\usepackage{amsmath,amssymb}
\usepackage{mathrsfs}

\usepackage{color}
\usepackage{here}

\begin{document}


\title{Long time behavior of 
nonlinear electromagnetic wave in vacuum 
beyond linear approximation}


\author{Kazunori Shibata}
\affiliation{Institute of Laser Engineering, Osaka University, 2-6 Yamada-Oka, Suita, Osaka 565-0871 Japan}


\date{\today}

\begin{abstract}
A quantum nature of vacuum is expected to affect 
electromagnetic fields in vacuum as a nonlinear 
correction, yielding nonlinear Maxwell's equations. 
We extend the finite-difference time-domain (FDTD) method 
in the case that the nonlinear electromagnetic Lagrangian is 
quartic with respect to the electric field and magnetic flux density. 
With this extension, the nonlinear Maxwell's equations 
can be numerically solved without making any assumptions 
on the electromagnetic field. 
We demonstrate examples of self-modulations of 
nonlinear electromagnetic waves in a one-dimensional cavity, 
in particular, in a time scale beyond an applicable 
range of linear approximation. 
A momentarily small nonlinear correction can accumulate and 
a comparably large self-modulation can be  
achieved in a long time scale 
even though the electromagnetic field is not extremely strong. 
Further, we analytically approximate the nonlinear 
electromagnetic waves in the cavity and clarify the characteristics, 
for example, how an external magnetic flux density changes 
the self-modulations of phase and polarization. 
\end{abstract}


\maketitle

\section{Introduction}

In classical electromagnetism, the electromagnetic fields 
in vacuum are 
described by the linear Maxwell's equations. In modern physics, 
several corrections have been proposed 
for the behavior of the 
electromagnetic fields such as the 
Heisenberg-Euler theory \cite{Heisenberg1936} 
based on the quantum electrodynamics, 
the Born-Infeld theory \cite{doi:10.1098-rspa.1934.0059} 
derived by an analogy to 
the special theory of relativity, and 
the more generalized 
Pleba{\'{n}}ski class \cite{Plebanski1970}. 
The nonlinear correction of electromagnetic fields 
affects many branches of physics. 
For example, several calculations are performed for 
the interaction of strong laser beams 
\cite{RevModPhys.84.1177}\cite{king_heinzl_2016}, 
the radiation from pulsars and neutron stars \cite{Heyl_2005}
\cite{Shakeri_2017}\cite{10.1093/mnras/stw2798}, 
the Wichmann-Kroll correction \cite{PhysRev.101.843} 
to the Lamb shift, 
a photon-photon scattering \cite{PhysRevAccelBeams.20.043402}, 
an interaction between a nucleus and 
electrons through 
the Uehling potential 
\cite{PhysRev.48.55}\cite{Frolov2012}\cite{FROLOV2014499}, 
a correction to the states of a hydrogen atom 
\cite{Akmansoy2018}\cite{PhysRevLett.96.030402}
\cite{Mazharimousavi2012}\cite{Denisov2006}, 
an electromagnetic effect for 
black holes \cite{AYONBEATO199925}\cite{PhysRevD.63.044005}, 
and a possibility of 
magnetic monopoles \cite{PhysRevD.63.044005}.

Various experimental proposals have been considered 
for the verification of the nonlinear correction, such as 
an inverse Cotton-Mouton effect \cite{Rizzo_2010}, 
four wave mixing \cite{PhysRevLett.96.083602}
\cite{PhysRevA.74.043821}, 
a refraction of light by light \cite{Sarazin2016}, and 
birefringence \cite{PintoDaSouza2006}
\cite{PhysRevD.93.093020}
\cite{PhysRevD.95.099902}\cite{Battesti_2012}. 
Several experiments have also been performed 
\cite{Bernard2000}\cite{DellaValle2016}\cite{Fan2017}
\cite{Cadene2014}, 
but the nonlinear correction has yet been observed. 
In many proposals and numerical evaluations, 
a nonlinear effect is calculated via 
a linear approximation of the nonlinear Maxwell's equations, 
{\it i.e.}, a large classical input induces the polarization and 
magnetization of vacuum and they act as a wave source 
for another relatively small electromagnetic wave 
\cite{PhysRevLett.96.083602}\cite{particles3010005}. 
In several previous studies, the nonlinearity, 
{\it i.e.}, the self-interaction, is 
partially included as nonlinear Schr\"{o}dinger equations via 
a slowly varying envelope approximation and 
a perturbation that a weak field propagates in a 
relatively large background field
\cite{RevModPhys.78.591}\cite{Rozanov1998}.

In this study, 
we explain a numerical method for solving the nonlinear 
Maxwell's equations without any 
assumptions on the electromagnetic field, 
{\it i.e.}, the nonlinear correction does not 
need to be comparably small and we do not have to 
assume an envelope function. 
To this end, an extension of the 
finite-difference time-domain (FDTD) 
method \cite{FDTD_Book} is given, followed by 
several examples of self-modulation 
in a one-dimensional cavity. 
Further, we analytically approximate the leading parts 
of the nonlinear electromagnetic waves 
to reproduce the numerical results. 
These demonstrations reveal an accurate 
time evolution of the resonant increase 
\cite{Shibata2020}\cite{Shibata2021EPJD}
\cite{PhysRevLett.87.171801}, which has been 
calculated by the linear approximation. 
Throughout the extension of the FDTD method and 
demonstrations in a one-dimensional cavity, 
we reveal a novel nonlinear property that 
a large nonlinear effect can appear by 
accumulating a momentarily small nonlinear correction 
for a long time scale, even though an input 
electromagnetic field is not extremely strong. 
For example, we demonstrate that the polarization 
can change by 90 degrees at a specific time 
if an adequate magnetic flux density is imposed.

\section{Basic notations}

We normalize the electromagnetic fields by 
the electric constant $\varepsilon_0$ and 
magnetic constant $\mu_0$. The  
electric field is multiplied by $\varepsilon_0^{1/2}$ and 
electric flux density is divided by $\varepsilon_0^{1/2}$. 
Similarly, the magnetic flux density 
is divided by $\mu_0^{1/2}$ and 
magnetic field is multiplied by $\mu_0^{1/2}$. 
Using the electric field $\boldsymbol{E}$ 
and magnetic flux density $\boldsymbol{B}$, we introduce 
two Lorentz invariants by $F=E^2-B^2$ 
and $G=\boldsymbol{E}\cdot\boldsymbol{B}$. 
The Lagrangian density we treat in this study is given by 
\begin{equation}\label{def_L}
\mathscr{L}=\frac{1}{2}F+C_{2,0}F^2+C_{0,2}G^2,
\end{equation}
where $C_{2,0}$ and $C_{0,2}$ are the 
nonlinear parameters\cite{PhysRevD.93.093020}. 
This Lagrangian is quartic with respect to the electric field and 
magnetic flux density. 
This form is frequently considered 
and regarded as an effective Lagrangian. 
In Figs. 2,3, and 4, we use the values 
$C_{2,0}=1.665\times 10^{-30}$(m$^3$/J) and $C_{0,2}=7C_{2,0}$ 
\cite{PhysRevD.93.093020}\cite{PhysRev.82.664} 
of the Heisenberg-Euler model. 
A part of electromagnetic field can be calculated by 
the classical linear Maxwell's equations. The 
``classical term'' is expressed by a subscript $c$. 
The difference from the classical term is the 
``corrective term'' and expressed by a subscript $n$. 
Thus, we can express as 
$\boldsymbol{E}=\boldsymbol{E}_c+\boldsymbol{E}_n$ 
and 
$\boldsymbol{B}=\boldsymbol{B}_c+\boldsymbol{B}_n$, 
respectively. 
The corrective electric flux density and magnetic field 
are given by 
\begin{subequations}\label{def_DH}
\begin{equation}\label{D_BE} 
\boldsymbol{D}_n=
\boldsymbol{E}_n+4C_{2,0}F\boldsymbol{E}
+2C_{0,2}G\boldsymbol{B}, 
\end{equation}
\begin{equation}\label{H_BE} 
\boldsymbol{H}_n=
\boldsymbol{B}_n+4C_{2,0}F\boldsymbol{B}
-2C_{0,2}G\boldsymbol{E}. 
\end{equation}
\end{subequations}
The nonlinear Maxwell's equations in vacuum 
for the corrective term are given by 
\begin{equation}\label{NME_hosei}
\begin{split}
&\nabla\cdot\boldsymbol{B}_n=0, \\
&\nabla\cdot\boldsymbol{D}_n=0, \\
&\partial_t \boldsymbol{B}_n=
-c\nabla\times\boldsymbol{E}_n,\\
&\partial_t \boldsymbol{D}_n=c\nabla\times\boldsymbol{H}_n,\\
\end{split}
\end{equation}
where $c$ is the speed of light 
and $\partial_t$ expresses the partial differentiation 
with respect to time $t$. 
The classical term can be numerically calculated 
or sometimes explicitly given in an analytic form 
without difficulties 
in the range of classical electromagnetism. 
Thus, the remaining 
problem is to solve the corresponding corrective term.

\section{Extension of FDTD method}

In many previous studies, 
analyses of linearized Maxwell's equations or 
nonlinear Schr\"{o}dinger equations have widely been performed 
with applying several approximations such as the corrective 
term is always much smaller than the classical term. 
We here explain an extension of the FDTD method which 
enables us to execute a numerical calculation without 
these approximations. 
As the name indicates, in the FDTD method, 
a time-evolution of electromagnetic fields is numerically 
calculated by a finite-difference method in the time-domain. 
Once the electromagnetic fields 
at a time step are given, the magnetic flux density and 
electric flux density at the next time step are numerically 
calculated by the discretized Maxwell's equations. 
Then, the electric field and magnetic field are calculated. 
The FDTD method can be executed straightforward 
in classical electromagnetism, as we can directly 
obtain the electric and magnetic fields because 
they are proportional to 
the electric and magnetic flux densities, respectively. 
On the contrary, in nonlinear electromagnetism, 
the electric field and electric flux density 
are not proportional and 
$\boldsymbol{E}_n$ is only implicitly given to 
satisfy Eq. (\ref{D_BE}) for numerically obtained 
$\boldsymbol{B}_n$ and $\boldsymbol{D}_n$. 
Thus, a special procedure 
for calculating $\boldsymbol{E}_n$ 
is required to execute the FDTD method for the 
next time step. Once $\boldsymbol{E}_n$ is obtained, 
$\boldsymbol{H}_n$ is directly obtained by Eq. (\ref{H_BE}) 
and we can proceed to the next time step.

\begin{figure}
\includegraphics[width=8.6cm]{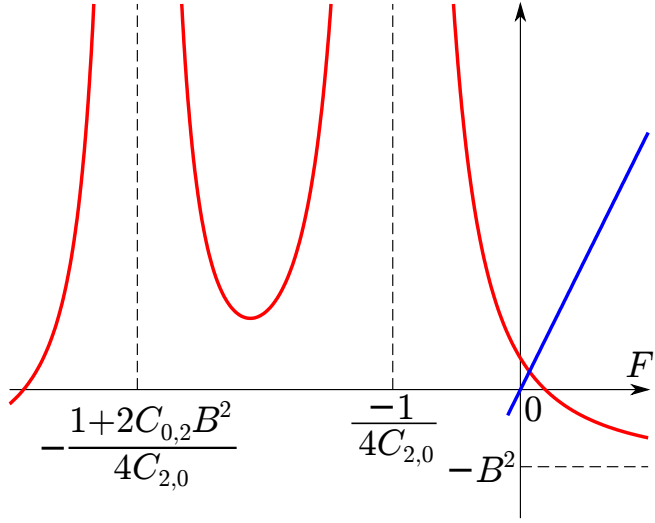} 
\caption{
Typical distributions 
of both sides of Eq. (\ref{F_kettei}) 
in the case of $D_1\neq0$ and $D_2\neq0$. 
The red curves are the left-hand side and 
the blue line is the right-hand side. 
The vertical axis is the value of each side. 
The unique intersection 
expresses the unique solution of $F$. 
The intersection exists at $F<0$ if $E^2<B^2$. 
\label{zu_F_kettei}}
\end{figure}

We explain the procedure. 
First, we obtain $\boldsymbol{B}$ and $\boldsymbol{D}$ 
by adding numerically obtained 
$\boldsymbol{B}_n$ and $\boldsymbol{D}_n$ to 
the given (or already calculated) 
$\boldsymbol{B}_c$ and $\boldsymbol{D}_c$, {\it i.e.}, 
$\boldsymbol{B}=\boldsymbol{B}_c+\boldsymbol{B}_n$ and 
$\boldsymbol{D}=\boldsymbol{D}_c+\boldsymbol{D}_n$. 
If $B\neq0$, let 
$D_1=\boldsymbol{D}\cdot\boldsymbol{B}/B$ 
and $D_2=|\boldsymbol{D}-D_1\boldsymbol{B}/B|$, 
we obtain
\begin{equation}\label{F_kettei}
\frac{D_1^2}{(1+4C_{2,0}F+2C_{0,2}B^2)^2}
+\frac{D_2^2}{(1+4C_{2,0}F)^2}-B^2
=F.
\end{equation}
If $B=0$, we can use $D^2$ instead of $D_1^2+D_2^2$. 
Since $\boldsymbol{B}$ and $\boldsymbol{D}$ are 
already calculated, this equation can be used to 
determine $F$. It is worth emphasizing that we can 
calculate $F$ even though we have yet obtained the 
electric field. 
Figure \ref{zu_F_kettei} shows both sides as 
functions of $F$. 
In the case of $D_2\neq0$, the 
left-hand side monotonically decreases 
at $F>-1/(4C_{2,0})$ and converges to $-B^2\le0$. 
The right-hand side obviously increases monotonically. 
Therefore, 
if $B^2<1/(4C_{2,0})$, 
we can obtain a unique $F$ that satisfies Eq. (\ref{F_kettei}) 
in the domain of $F>-1/(4C_{2,0})$. 
Then, we can calculate 
a matrix $\Lambda$ and a vector $\boldsymbol{\chi}$ 
by 
\begin{equation}\label{}
\begin{split}
&\Lambda=(1+4C_{2,0}F)
\begin{pmatrix}
1 & 0 & 0 \\
0 & 1 & 0 \\
0 & 0 & 1 \\
\end{pmatrix}
+2C_{0,2}
\begin{pmatrix}
B_x^2 & B_xB_y & B_xB_z \\
B_xB_y & B_y^2 & B_yB_z \\
B_xB_z & B_yB_z & B_z^2 \\
\end{pmatrix},\\
&\boldsymbol{\chi}=\boldsymbol{D}_n
-4C_{2,0}F\boldsymbol{E}_c-2C_{0,2}
(\boldsymbol{E}_c\cdot\boldsymbol{B})\boldsymbol{B},\\
\end{split}
\end{equation}
as independent values of $\boldsymbol{E}_n$. 
Because $|\Lambda|\neq0$, 
$\boldsymbol{E}_n$ is uniquely obtained by 
\begin{equation}\label{}
\boldsymbol{E}_n=\Lambda^{-1}\boldsymbol{\chi}.
\end{equation}
In the case of $D_1\neq0$ and $D_2=0$, unique 
$F$ and $\boldsymbol{E}_n$ are obtained in a similar way. 
If $D_1=0$ and $D_2=0$, $\boldsymbol{E}_n=-\boldsymbol{E}_c$ 
is clear. Then, the procedure is established.

\section{Examples in one-dimensional cavity}

We demonstrate numerical calculations 
in a one-dimensional cavity system 
with length $L$
in the $x$ direction, {\it i.e.}, $0\le x\le L$. 
The mirrors are supposed to be perfect conductors 
and the boundary conditions are given as the 
$y,z$ components of the electric field and 
the $x$ component of the magnetic flux density 
to be zero. 
The classical term at $t\ge 0$ is given as 
the sum of a standing wave and static magnetic flux density:  
\begin{equation}\label{tei_EB_with_sta}
\begin{split}
&\boldsymbol{E}_c=A \sin \omega t \sin kx \boldsymbol{e}_y, \\
&\boldsymbol{B}_c=A \cos \omega t \cos kx\boldsymbol{e}_z+
\boldsymbol{B}_s, \\
\end{split}
\end{equation}
where $A$ is the amplitude of the standing wave, 
the wave number $k$ and frequency $\omega$ are 
connected to the wave length $\lambda$ via 
$k=2\pi/\lambda$ and $\omega=ck$, and 
$\boldsymbol{e}_{y,z}$ are the unit vectors 
of the $y,z$ directions, respectively. 
We employ $\lambda=400$ (nm) and 
$L=100\pi/k$ for numerical calculations. 
$\boldsymbol{B}_s=(0,B_{sy},B_{sz})$ 
is a constant static magnetic flux density 
and its magnitude is expressed by $B_s$. 
This external field is highly valuable 
because it can vary the behavior of 
the nonlinear correction. 
For this system and classical term, the $x$ component 
of all fields are always zero and we do not mention hereinafter. 
We suppose $(C_{2,0}+C_{0,2})(A^2+B_s^2)\ll 1$ 
because the nonlinear Lagrangian in Eq. (\ref{def_L}) is 
limited to be quartic. 
The condition of $B^2<1/(4C_{2,0})$ always holds for the 
present calculation. 
For the given classical term, 
we calculate the corrective term at $t\ge 0$. 
The initial values of both corrective electric field and 
magnetic flux density are set to be zero everywhere 
because they should be much smaller than $A$.

\begin{figure}
\includegraphics[width=8.6cm]{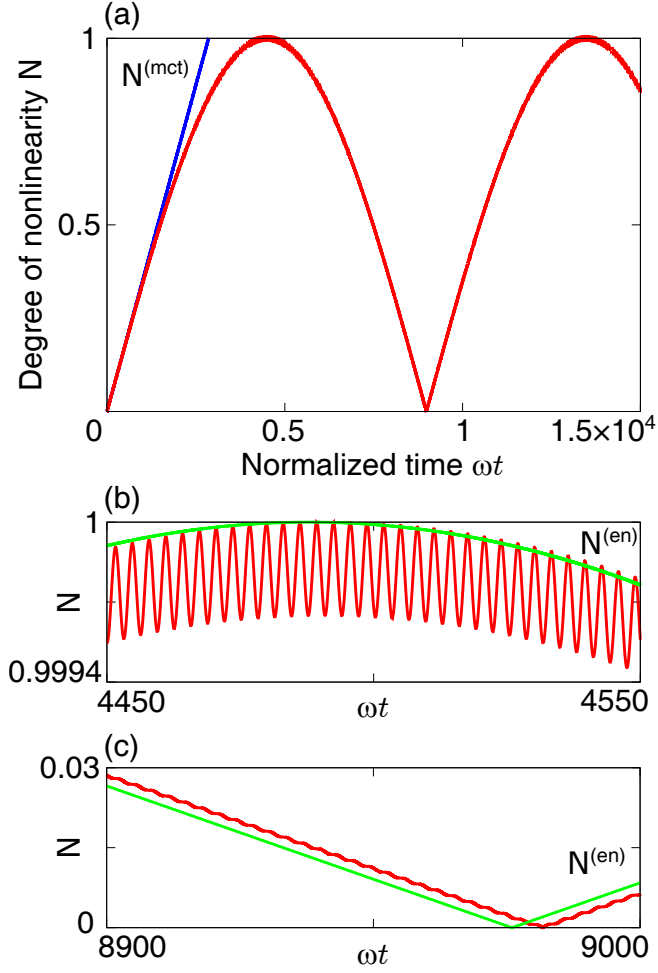} 
\caption{
The degree of nonlinearity $\mathsf{N}$ for 
$A=10^{-6}/\sqrt{C_{2,0}},B_{sy}=10^{-2}/\sqrt{C_{2,0}}$, 
and $B_{sz}=0$. 
In (a), $\mathsf{N}^{(\text{mct})}$ is also shown. 
(b,c) are enlarged graphs at the first peak and 
zero, respectively. 
The leading part $\mathsf{N}^{(\text{lp})}$ in Eq. (\ref{ol_fig2})
is also shown. 
\label{A-6Bsy-2Bsz0}}
\end{figure}
\begin{figure}
\includegraphics[width=8.6cm]{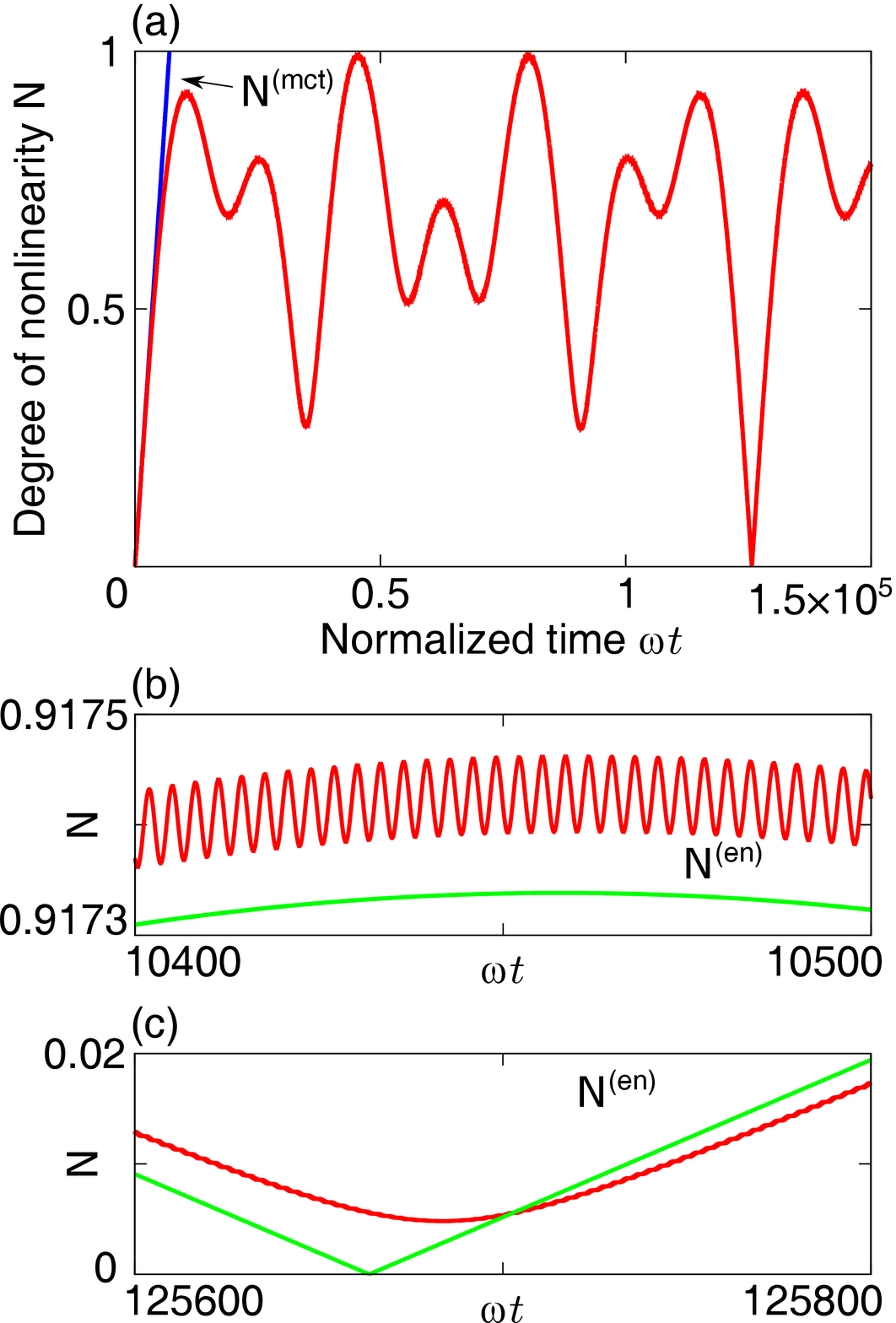} 
\caption{
The degree of nonlinearity $\mathsf{N}$ for 
$A=10^{-6}/\sqrt{C_{2,0}}$ 
and $B_{sy}=B_{sz}=0.5\times 10^{-2}/\sqrt{C_{2,0}}$. 
In (a), $\mathsf{N}^{(\text{mct})}$ is also shown. 
(b,c) are enlarged graphs at the first peak and 
zero, respectively. 
The leading part $\mathsf{N}^{(\text{lp})}$ in Eq. (\ref{ol_fig3})
is also shown. 
\label{A-6ByBz-2waru2}}
\end{figure}

If $A$ and $B_s$ are 
too small, a huge calculation time is required to see the 
nonlinear effect beyond the linear approximation. 
Because of this numerical limitation, 
we first calculate with unrealistic large parameters. 
An evaluation for realistic values is performed later in 
Fig. \ref{Pol_daen}. 
The amplitude of the standing wave 
is set to $A=10^{-6}/\sqrt{C_{2,0}}$ 
and the corresponding intensity $cA^2$ is about 
$1.80\times 10^{22}$ (W/cm$^2$). 
For the static magnetic flux density, 
$B_{sy}=10^{-2}/\sqrt{C_{2,0}}$ and $B_{sz}=0$ are used in 
Fig. \ref{A-6Bsy-2Bsz0} and 
$B_{sy}=B_{sz}=0.5\times 10^{-2}/\sqrt{C_{2,0}}$ are 
used in Fig. \ref{A-6ByBz-2waru2}. 
The value $10^{-2}/\sqrt{C_{2,0}}$ corresponds to 
$10^{-2}\sqrt{\mu_0/C_{2,0}} \approx 8.7 \times 10^9$ (T). 
While these are too large for experiments 
in a cavity, the calculation 
itself is consistent because 
$(C_{2,0}+C_{0,2})(A^2+B_s^2)\ll 1$ holds.

To visualize the nonlinear effect, 
we define a temporal 
function that expresses the magnitude 
of the corrective term throughout the cavity. 
Let 
\begin{equation}
E_{ny}^{(\text{sq})}(t)=\max_{0\le x\le L}E_{ny}^2(x,t),
\end{equation}
and defining $E_{nz}^{(\text{sq})}(t)$, 
$B_{ny}^{(\text{sq})}(t)$, and $B_{nz}^{(\text{sq})}(t)$ 
in a similar way, 
we introduce ``the degree of nonlinearity'' as 
\begin{equation}\label{deg_Nonl_N}
\mathsf{N}(t)=\frac{1}{2A}
\left[
E_{ny}^{(\text{sq})}(t)+E_{nz}^{(\text{sq})}(t)
+B_{ny}^{(\text{sq})}(t)+B_{nz}^{(\text{sq})}(t)
\right]^{\frac{1}{2}},
\end{equation}
to indicate the strength of the nonlinear effect. 
$\mathsf{N}$ expresses a magnitude ratio of the 
corrective term to the classical term.  
For example, if $\mathsf{N}$ is much smaller than unity, 
a linear approximation will be applicable. 
On the contrary, when $\mathsf{N}$ is comparable to unity, 
such an electromagnetic field will not be analyzed by 
the linear approximation.

We first pay attention to the short time scale.  
Figures \ref{A-6Bsy-2Bsz0}(a) and \ref{A-6ByBz-2waru2}(a) 
show that $\mathsf{N}$ increases almost linearly. 
In this time scale, 
the nonlinear correction can be 
calculated with a linear approximation. 
The corresponding correction of electromagnetic fields 
were called as ``minimum corrective term'' in 
Ref. \cite{Shibata2021EPJD} 
and it shows the resonant increase. 
Let $\mathsf{N}^{(\text{mct})}$ be 
the degree of nonlinearity for the 
minimum corrective term. 
According to Eq. (18) in Ref. \cite{Shibata2021EPJD}, 
$\mathsf{N}^{(\text{mct})}$ 
behaves almost linear and is consistent 
with $\mathsf{N}$ in the short time scale, as in 
Figs. \ref{A-6Bsy-2Bsz0}(a) and \ref{A-6ByBz-2waru2}(a). 
As for a longer time scale, 
the linear approximation becomes an overestimation and 
$\mathsf{N}$ departs from $\mathsf{N}^{(\text{mct})}$.

Both Figs. \ref{A-6Bsy-2Bsz0} and \ref{A-6ByBz-2waru2} 
indicate that 
$\mathsf{N}$ has a slowly varying component 
as in both panels (a), as well as an oscillation 
with a period of about $2\pi/\omega$, 
as in panels (b) and (c). 
The slower variation is clearly a 
characteristic of nonlinear electromagnetic waves. 
In addition, 
Fig. \ref{A-6ByBz-2waru2}(a) shows 
more complicated behavior than 
Fig. \ref{A-6Bsy-2Bsz0}(a). 
This may arise from the 
energy transfer between two polarization modes 
through $B_{sy}$ and $B_{sz}$.

Readers may wonder how the demonstrated 
results are related to various previous calculations. 
The extension of FDTD method is done 
without making any assumptions 
except for the form of the Lagrangian. 
Therefore, it is possible to reproduce the calculation 
results obtained by a linear approximation or a 
nonlinear Schr\"{o}dinger equation. Furthermore, 
the present scheme is available for the outside 
of applicable range of these approximations, 
in particular, in a time scale when the corrective term 
becomes comparable to the classical term.

\section{Approximation by leading-part functions}

We further inquire into the demonstrated 
example and reveal a nonlinear characteristic mathematically. 
The numerical results indicate that 
the spatial distributions of the 
electric field are almost always 
proportional to $\sin kx$ and 
high-harmonic components are vanishing. 
These distributions might be attributed to the 
resonant behavior in the linear approximation: 
where the resonant increase is 
proportional to $\omega t\cos\omega t\sin kx$
\cite{Shibata2021EPJD}. 
We can expect then that 
the leading part of the electric field will be 
approximated by 
a product of a temporal function and $\sin kx$. 
By expressing by a superscript $(\text{lp})$, 
the leading part of the electric field and magnetic flux density 
can be supposed to be 
\begin{equation}\label{approx_EnBn_main}
\begin{split}
&E_y^{(\text{lp})}(x,t)
=[f\cos\omega t+g\sin \omega t]\sin kx,\\
&E_z^{(\text{lp})}(x,t)
=(l\cos\omega t+m\sin \omega t)\sin kx,\\
&B_y^{(\text{lp})}(x,t)
=(-m\cos\omega t+l\sin \omega t)\cos kx+B_{sy},\\
&B_z^{(\text{lp})}(x,t)
=[g\cos\omega t-f\sin \omega t]\cos kx+B_{sz},\\
\end{split}
\end{equation}
where $f,g,l,$ and $m$ depend only on time. They vary 
relatively slower than $\sin \omega t$ and $\cos\omega t$, 
{\it i.e.}, they can be regarded as slowly varying envelopes. 
They are determined 
by the following nonlinear differential equations: 
\begin{equation}\label{DE_hourakusen_main}
\begin{pmatrix}
f'\\
g'\\
l'\\
m'\\
\end{pmatrix}
=
\begin{pmatrix}
0&
-\mathscr{X}_1-al^2&
0&
-\xi+afl\\
\mathscr{X}_1+am^2&
0&
\xi-agm&
0\\
0&
-\xi+afl&
0&
-\mathscr{X}_2-af^2\\
\xi-agm&
0&
\mathscr{X}_2+ag^2&
0\\
\end{pmatrix}
\begin{pmatrix}
f\\
g\\
l\\
m\\
\end{pmatrix}, 
\end{equation}
where 
\begin{equation}
\begin{split}
&X=f^2+g^2+l^2+m^2,\\
&\mathscr{X}_1=4C_{2,0}B_{sz}^2+C_{0,2}B_{sy}^2+2C_{2,0}X,\\
&\mathscr{X}_2=4C_{2,0}B_{sy}^2+C_{0,2}B_{sz}^2+2C_{2,0}X,\\
&\xi=-(4C_{2,0}-C_{0,2})B_{sy}B_{sz},\\
&a=-2C_{2,0}+\frac{1}{2}C_{0,2}.\\
\end{split}
\end{equation}
A detailed derivation of the differential equations and 
their solutions are given in Appendix. 
It can be easily verified that $X=A^2$ is a conservative quantity, 
Note that this fact expresses a physical meaning that 
a leading part of the total energy conserves in the 
form of Eq. (\ref{approx_EnBn_main}) and 
the time evolution in Eq. (\ref{DE_hourakusen_main}).

In this approximation, the corrective term can be 
approximated as 
$\boldsymbol{E}_n \approx 
\boldsymbol{E}^{(\text{lp})}-\boldsymbol{E}_c$ and 
$\boldsymbol{B}_n \approx 
\boldsymbol{B}^{(\text{lp})}-\boldsymbol{B}_c$. Therefore, 
the leading part of the degree of nonlinearity 
$\mathsf{N}^{(\text{lp})}$ can be expressed by 
\begin{equation}\label{moto_A28_Nlp}
\mathsf{N}^{(\text{lp})}(t)
=\sqrt{\frac{1}{2}\left(1-\frac{g}{A}\right)}. 
\end{equation}

We write down the leading part of the 
total electric field 
$\boldsymbol{E}^{(\text{lp})}$ and 
$\mathsf{N}^{(\text{lp})}$ for both Figs. 
\ref{A-6Bsy-2Bsz0} and \ref{A-6ByBz-2waru2}. 
For Fig. \ref{A-6Bsy-2Bsz0}, Eq. (A.21) gives 
\begin{equation}\label{ol_fig2}
\begin{split}
&E_y^{(\text{lp})}
=A\sin (1-\mathscr{X}_1) \omega t \sin kx,\\
&E_z^{(\text{lp})}=0,\\
&\mathsf{N}^{(\text{lp})}(t)
=\left|\sin\frac{\mathscr{X}_1\omega t}{2}\right|,\\
\end{split}
\end{equation}
where $\mathscr{X}_1=C_{2,0}(2A^2+7B_s^2)$. 
We can see that $\mathsf{N}^{(\text{lp})}$ becomes zero when 
$\omega t$ is an integer multiple of 
$2\pi/\mathscr{X}_1 \approx 8.976 \times 10^3$, 
in accordance with the numerical result 
as in Fig. \ref{A-6Bsy-2Bsz0}(c). 
Equations (A.22) and (A.23) 
are used for Fig. \ref{A-6ByBz-2waru2}. 
In the time scale of $C_{2,0}A^2\omega t \ll 1$, 
we obtain 
\begin{equation}\label{ol_fig3}
\begin{split}
&E_y^{(\text{lp})}
\approx A\cos\xi \omega t 
\sin (1-11\xi/3)\omega t \sin kx,\\
&E_z^{(\text{lp})}
\approx -A\sin\xi \omega t 
\cos (1-11\xi/3)\omega t \sin kx,\\
&\mathsf{N}^{(\text{lp})}(t)
\approx
\sqrt{\frac{1}{2}\left(1-\cos\xi \omega t 
\cos\frac{11}{3}\xi \omega t \right)},\\
\end{split}
\end{equation}
where $\xi=(3/2)C_{2,0}B_s^2$. 
In this case, $\mathsf{N}^{(\text{lp})}$ 
becomes zero when 
$\omega t$ is an integer multiple of 
$2\pi/(C_{2,0}B_s^2)=
4\pi \times 10^4 \approx 1.2566 \times 10^5$, 
reproducing the numerical result 
as in Fig. \ref{A-6ByBz-2waru2}(c). 

We can see that $\mathsf{N}^{(\text{lp})}$ does not 
oscillate rapidly but 
well reproduces a rough behavior of 
$\mathsf{N}$: 
the difference is almost indistinguishable 
in the scale of panel (a) of both figures. 
Thus, it is shown only in panels (b) and (c). 
The slight differences 
are attributed to the discarded terms in 
the analytic calculation. 
Furthermore, $\boldsymbol{E}_n$ and 
$\boldsymbol{B}_n$ in a short time scale agree with 
the resonant increase of the minimum corrective term 
\cite{Shibata2021EPJD}, as calculated in Eq. (A.25). 
These results confirm the 
validity of the approximation of leading part 
in the present time scale. 
Note that it is not clear that the 
approximation is valid for an infinitely long time scale. 

These results suggest that 
the nonlinear effect in the one-dimensional cavity 
appears as changes of the phase and polarization. 
In contrast, a change in wavelength or frequency 
must be discrete 
because of the fixed boundary conditions and 
high-harmonic components are scarcely generated 
in the viewpoint of energy conservation.

In the present system, 
the maximum of $\mathsf{N}$ is about unity. 
It can be understood from 
Eq. (\ref{moto_A28_Nlp}) as it shows $\mathsf{N}^{(\text{lp})}\le 1$. 
When $\mathsf{N}^{(\text{lp})}\approx 1$, 
we can see that 
$E_y^{(\text{lp})}\approx -A\sin \omega t \sin kx $ is 
necessary, {\it i.e.}, the whole nonlinear 
electromagnetic wave 
is exactly the antiphase to the classical electromagnetic wave, 
and the phase shift becomes maximum. 
In Fig. \ref{A-6Bsy-2Bsz0}, 
$\mathsf{N}^{(\text{lp})}=1$ 
is realized when $\omega t$ is an 
odd integer multiple of 
$\pi/\mathscr{X}_1 \approx 4.488 \times 10^3$, 
as in Fig. \ref{A-6Bsy-2Bsz0}(b).

\section{Calculation for realistic parameters}

The leading-part functions enable us to clarify 
the behavior of nonlinear electromagnetic waves 
in much longer time scale than the extended FDTD method 
can reach. In particular, the leading-part functions are 
highly useful in the case that the classical amplitude $A$ 
and the magnitude of the external magnetic flux density $B_s$ 
are relatively small. If we try to run a numerical calculation 
with the extended FDTD method 
up to a time scale when the nonlinearity becomes dominant, 
an unrealistic long calculation time will be required.

We perform a realistic calculation of 
the classical amplitude 
to be $1.94\times10^6$ (V/m), corresponding to 
$10^6$ (W/cm$^2$) \cite{PhysRevA.91.052503} 
and the static magnetic flux density 
to be $5\sqrt{2}$ (T) for both $y$ and $z$ components 
\cite{Durrell_2014}\cite{Durrell_2018}. 
Because $A$ and $\boldsymbol{B}_s$ are realistic, {\it i.e.},  
much smaller than the above values, 
the leading-part functions will be 
sufficiently precise approximations. 

\begin{figure}
\includegraphics[width=8.6cm]{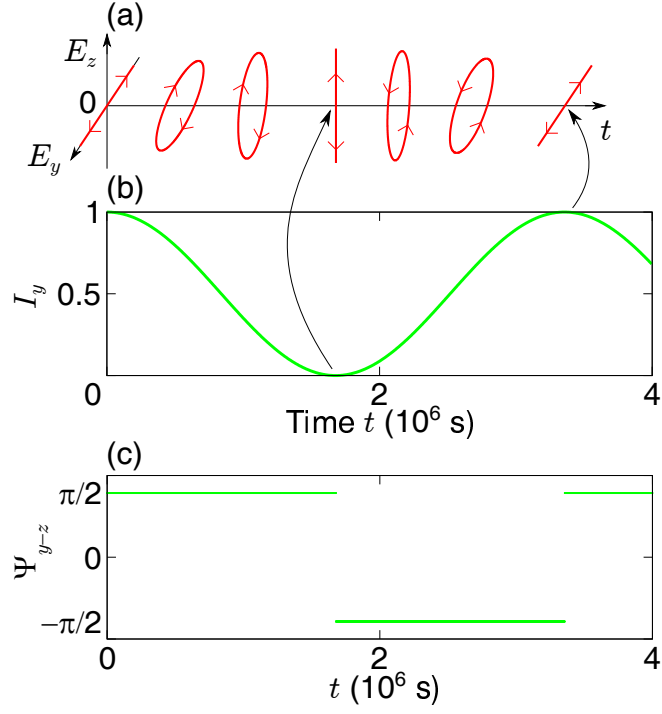}
\caption{
The time evolution of the polarization 
at a position of $\sin kx=1$. (a) The shapes and rotation 
directions of the elliptical polarization. 
The axes of ellipses are almost 
coincident with the $y$ and $z$ axes. 
(b) The intensity 
ratio $I_y$ in Eq. (\ref{wariai_yz}). (c) The relative phase 
$\Psi_{y-z}$ in Eq. (A.29). Its sign determines the 
rotation direction. 
\label{Pol_daen}}
\end{figure}

We demonstrate the time evolution of the polarization. 
For this purpose, we calculate the intensity ratio of 
the $y$ component of the electric field 
to the total electric field and the 
relative phase. 
Using Eq. (A.26), the intensity ratio $I_y$ is given as 
\begin{equation}\label{wariai_yz}
I_y=\frac{1}{2}
[1+\text{cn}(p\omega t,iq)\text{dn}(p\omega t,iq)], 
\end{equation}
where $p$ and $q$ are given in Eq. (A.17). 
The relative phase $\Psi_{y-z}$ is defined in Eq. (A.29). 
Figure \ref{Pol_daen} shows a result at a fixed 
point of $\sin kx=1$. 
Figure \ref{Pol_daen}(a) is a typical time evolution of the 
polarization mode. It varies between 
two orthogonal linear polarizations. 
During the transition, the polarization is 
almost elliptic because the magnitude and phase of 
each component of electric field scarcely change in a 
cycle of $2\pi/\omega$. 
$I_y$ and $\Psi_{y-z}$ are shown in 
Figs. \ref{Pol_daen}(b) and \ref{Pol_daen}(c), respectively. 
The value of 
$I_y$, well approximated by $\cos^2\xi \omega t$, 
determines the shape of the ellipse of polarization 
and the sign of $\Psi_{y-z}$ determines the rotation direction. 
In particular, the polarization changes by 
90 degrees from $y$ to $z$ at 
$t\approx \pi/(2\xi \omega)
=\pi/(3C_{2,0}B_s^2\omega)\approx1.68 \times 10^6$ 
seconds.

We briefly discuss an experimental perspective for the 
demonstrated example. 
Suppose we connect the ratio $I_y$ and 
a detectable polarization angle $\theta$ (deg) 
of precise measurement, we can estimate a necessary 
time to confine the standing wave in the cavity by 
$\cos^2\xi \omega t\approx \cos(\theta\pi/180)$, 
yielding 
\begin{equation}\label{}
t\approx \frac{\mu_0\lambda}{3\pi cC_{2,0}B_s^2}
\text{Cos}^{-1}\sqrt{\cos\frac{\theta}{180}\pi}. 
\end{equation}
If we take $\theta=0.003$ \cite{Mukherjee2019}
and $B_{sy}=B_{sz}=30$ (T) \cite{Majkic_2020}, 
the necessary time is $t\approx 2.20$ seconds. 
On the other hand, 
even if a high reflectivity mirror is 
employed such as in gravitational wave detectors
\cite{PhysRevApplied.14.014021}\cite{10.1093/ptep/ptaa125}
\cite{coatings6040061}, 
a light in the cavity vanishes within several milliseconds 
for a cavity length of $L=30$ (cm). 
There is a gap of time scales by 3-digits. 
To bridge the gap or to find other features of 
self-modulation may be important open problems 
for future verification experiments. 
For example, to decrease the loss by reflection, 
a longer cavity with partial external fields 
or compensation by a successive adequate input will be reasonable. 
The extension of FDTD method is also useful for such calculations. 
In addition, realistic boundary conditions 
will be necessary for 
a high reflectivity mirror with a multilayer stack 
\cite{coatings6040061}\cite{Sidqi:19}.

\section{Final remarks}

We have extended the FDTD method to 
execute a numerical calculation 
without making any assumptions 
except for the form of the nonlinear Lagrangian to be quartic. 
We demonstrated the nonlinear electromagnetic waves 
in a one-dimensional cavity. We further derived an analytic 
approximation as the leading-part functions and 
mathematically clarified a characteristic of the 
nonlinear standing waves. 
The numerical and analytical results in a one-dimensional 
cavity are not simply as same as the well-known birefringence. 
In a calculation of the birefringence, 
a propagating wave in a cavity is frequently 
assumed to be sufficiently 
smaller than an external field and to be 
a plane-wave eigenmode, resulting in 
the time-independent 
dispersion relation \cite{PhysRevD.93.093020}\cite{Battesti_2012}
\cite{doi:10.1142S021974991241002X}\cite{Della_Valle_2013}. 
Several studies show a time-dependent dispersion relation 
by using a perturbation 
\cite{Battesti_2012}\cite{doi:10.1063/1.1759628}. 
The present calculations are performed without these assumptions. 
For example, the propagating wave does not have to be 
smaller than the external field, in particular, the phase 
can self-modulate without external field. 
We have not considered a dispersion relation because 
it is not adequate to approximate 
a nonlinear electromagnetic wave in a cavity by a plane wave.

The extended 
FDTD method is applicable to even more general systems, 
{\it i.e.}, not limited to a cavity system,  
and may reproduce numerous previous results obtained 
by a linear approximation or a 
nonlinear Schr\"{o}dinger equation. 
For example, if the time scale of nonlinear interaction is 
extremely short, such as a focusing of high power lasers, 
calculation results of the extended FDTD method and 
the linear approximation will be in good agreement. 
The most important physical picture 
of the present study is that 
a momentarily small nonlinear effect can accumulate and 
can appear as a comparably large self-modulation 
in a long time scale, 
even though 
the input or classical electromagnetic field is small. 
Such a novel property is not calculated if the corrective term 
is assumed to be always sufficiently smaller than the classical term. 
While, the larger electromagnetic fields are preferable because 
the nonlinear effect can appear in shorter time. 
The extended FDTD method 
will enable us to discover novel 
properties of nonlinear electromagnetic waves in a time scale 
when the nonlinearity becomes dominant, 
yielding new and optimized verification experiments 
of nonlinear electromagnetism in vacuum.

\appendix*

\renewcommand{\theequation}{A.\arabic{equation} }
\setcounter{equation}{0}

\subsection{Appendix: Calculation of 
the leading-part functions}

In this appendix, we explain a detailed analysis 
of the leading-part functions in Eq. (\ref{approx_EnBn_main}). 
The derivation of the differential equations 
in Eq. (\ref{DE_hourakusen_main}) and their solutions are shown. 
Originally, the leading-part functions of the 
electric field and magnetic flux density are introduced as 
\begin{equation}\label{approx_EnBn}
\begin{split}
&E_y^{(\text{lp})}(x,t)
=[f\cos\omega t+(g-A)\sin \omega t]\sin kx,\\
&E_z^{(\text{lp})}(x,t)
=(l\cos\omega t+m\sin \omega t)\sin kx,\\
&B_y^{(\text{lp})}(x,t)
=(-\tilde{m}\cos\omega t+\tilde{l}\sin \omega t)\cos kx+B_{sy},\\
&B_z^{(\text{lp})}(x,t)
=[(\tilde{g}-A)\cos\omega t-\tilde{f}\sin \omega t]\cos kx+B_{sz}.\\
\end{split}
\end{equation}
In the following calculations, 
$T=\omega t$ is used as a new variable. 
All the functions of $f,g,l,m$, and tilde-added ones 
depend only on $T$ and supposed to vary 
slower than $\sin T$ and $\cos T$. 
They are determined by 
the nonlinear Maxwell's equations given as 
\begin{equation}\label{1d_NME}
\begin{split}
&-\frac{\partial}{\partial x}E_z^{(\text{lp})}
+c^{-1}\frac{\partial}{\partial t}B_y^{(\text{lp})}=0,\\
&\frac{\partial}{\partial x}E_y^{(\text{lp})}
+c^{-1}\frac{\partial}{\partial t}B_z^{(\text{lp})}=0,\\
&-\frac{\partial}{\partial x}H_z^{(\text{lp})}
-c^{-1}\frac{\partial}{\partial t}D_y^{(\text{lp})}=0,\\
&\frac{\partial}{\partial x}H_y^{(\text{lp})}
-c^{-1}\frac{\partial}{\partial t}D_z^{(\text{lp})}=0.\\
\end{split}
\end{equation}

\subsubsection{1st and 2nd lines of Maxwell's equations}
The first line of the Maxwell's equations gives 
\begin{equation}\label{}
(l-\tilde{l}+\tilde{m}')\cos T+(m-\tilde{m}-\tilde{l}')\sin T=0. 
\end{equation}
Each parenthesis varies slowly and 
is supposed to be zero, {\it i.e.}, 
\begin{equation}\label{lm_EB}
l-\tilde{l}+\tilde{m}'=0, \ \ m-\tilde{m}-\tilde{l}'=0. 
\end{equation}
Similarly, the second line gives 
\begin{equation}\label{}
(f-\tilde{f}+\tilde{g}')\cos T+(g-\tilde{g}-\tilde{f}')\sin T=0, 
\end{equation}
yielding 
\begin{equation}\label{fg_EB}
f-\tilde{f}+\tilde{g}'=0, \ \ g-\tilde{g}-\tilde{f}'=0. 
\end{equation}

Because $f$ and the other functions vary relatively slowly, 
their derivatives will be comparably small if they are not 
identically zero. 
Combining with above suppositions, 
it will be adequate to suppose the following: 
\begin{equation}\label{katei_nyoro_onaji}
\begin{split}
&f \approx \tilde{f}, \ \ g \approx \tilde{g}, \ \ 
l \approx \tilde{l}, \ \ m \approx \tilde{m}, \\
& f' \approx \tilde{f}', \ \ g' \approx \tilde{g}', \ \ 
l' \approx \tilde{l}', \ \ m' \approx \tilde{m}'. \\
\end{split}
\end{equation}
Equations (A.4), (A.6), and (A.7) enable us to approximate as 
\begin{equation}\label{}
\begin{split}
&\tilde{f}=f+g',\\
&\tilde{g}=g-f',\\
&\tilde{l}=l+m',\\
&\tilde{m}=m-l',\\
\end{split}
\end{equation}
and we erase the four tilde-added functions from our calculation. 
In addition, the tilde-added functions are replaced 
in Eq. (\ref{approx_EnBn_main}) in the main text.

\subsubsection{3rd and 4th lines of Maxwell's equations}
Substituting the classical and corrective terms into 
the third and fourth lines of the Maxwell's equations 
shows a dependence on $x$ 
by $\sin kx$ and $\sin3kx$. We only retain 
the part of $\sin kx$. 
For the temporal part, the terms which 
oscillate faster than $\cos T$ and $\sin T$ are discarded, 
such as $\sin 2T$. 
Finally, the terms multiplied by $\cos T$ and $\sin T$ 
are regarded as zero independently, as in the above postulates, 
we obtain four differential equations on the 
slowly varying functions. 
Each of $f,g,l$, and $m$ is expected to be at most of the 
order of $A$. 
The products of a nonlinear parameter and two functions 
such as $C_{2,0}fl$ will be much smaller than unity. 
Therefore, the product of such values and 
the derivatives are excluded. 
Finally, the simultaneous differential equations 
in Eq. (\ref{DE_hourakusen_main}) are derived.

\subsection{Solution}
We solve the equations in Eq. (\ref{DE_hourakusen_main}) 
given the initial conditions of $f(0)=0, g(0)=A, l(0)=0$, 
and $m(0)=0$. 
First, we can see $X'=0$. Therefore, 
$X=A^2$ is constant and 
$\mathscr{X}_1$ and $\mathscr{X}_2$ are also constant. 
In particular, all of $f,g,l$, and $m$ are bounded and 
at most of the order of $A$. 
We introduce three variables by 
\begin{equation}
\alpha=f^2+g^2, \ \ 
\beta=fl+gm, \ \ 
\gamma=-fm+gl. 
\end{equation}
Defining a new constant as 
\begin{equation}
\Delta=\mathscr{X}_1-\mathscr{X}_2, 
\end{equation}
we obtain differential equations for 
$\alpha, \beta$, and $\gamma$ as 
\begin{equation}
\begin{split}
&\alpha'=2\gamma(\xi-a\beta),\\
&\beta'=\gamma[-\Delta+a(2\alpha-X)],\\
&\gamma'=\Delta\beta+\xi(X-2\alpha),\\
\end{split}
\end{equation}
with the initial conditions of 
$\alpha(0)=A^2,\beta(0)=0$, 
and $\gamma(0)=0$ . 
Because $f^2+g^2+l^2+m^2=A^2$, the range of 
$\alpha,\beta$, and $\gamma$ 
are bounded. Hence, the set of differential equations 
is Lipschitz continuous and the solution for the 
initial value problem is unique. 

We first calculate for $\xi\neq0$. Let 
\begin{equation}\label{}
Z=\Delta(-\Delta+aX)-4\xi^2,
\end{equation}
and also 
\begin{equation}\label{}
p=\sqrt{\frac{-Z+\sqrt{Z^2+4\xi^2a^2X^2}}{2}}, \ \ 
q=\frac{-\xi aX}{p^2}, 
\end{equation}
we obtain 
\begin{equation}\label{alphabetagamma_xineq0Deltaneq0}
\begin{split}
&\alpha=\frac{X}{4\xi^2+\Delta^2}\left\{
\Delta^2+\Delta\xi q\text{sn}^2(pT,iq)
+2\xi^2\left[1+\text{cn}(pT,iq)\text{dn}(pT,iq)\right]
\right\},\\
&\beta=\frac{\xi X}{4\xi^2+\Delta^2}\left\{
2\xi q\text{sn}^2(pT,iq)
+\Delta\left[1-\text{cn}(pT,iq)\text{dn}(pT,iq)\right]
\right\},\\
&\gamma=-\frac{\xi X}{p}\text{sn}(pT,iq).\\
\end{split}
\end{equation}
We give $f,g,l$, and $m$ for individual cases. 
In any case, their derivatives are sufficiently 
smaller than the maximum value of the original function 
and consistent with the discussion and assumptions around 
Eq. (A.7).

\subsubsection{$f,g,l$, and $m$ for 
$\xi\neq0$ and $\Delta\neq 0$}
The case of $\xi\neq0$, and $\Delta\neq 0$ 
corresponds to 
$B_{sy}\neq0, B_{sz}\neq0$, and $B_{sy}\neq B_{sz}$. 
In this case, $\alpha>0$ always holds and 
$\sqrt{\alpha}$ is always differentiable. 
Therefore, let 
\begin{equation}
\Theta=\mathscr{X}_1T
+\int_0^T\frac{\xi\beta(\tau)+a\gamma(\tau)^2}{\alpha(\tau)}\text{d}\tau, 
\end{equation}
we obtain 
\begin{equation}\label{fglm_xi_neq0_Delta_neq0}
\begin{split}
&f=-\sqrt{\alpha}\sin\Theta,\\
&g=\sqrt{\alpha}\cos\Theta,\\
&l=\frac{1}{\sqrt{\alpha}}
(-\beta\sin\Theta+\gamma\cos\Theta),\\
&m=\frac{1}{\sqrt{\alpha}}
(\beta\cos\Theta+\gamma\sin\Theta).\\
\end{split}
\end{equation}

\subsubsection{$f,g,l$, and $m$ for 
$\xi\neq0$ and $\Delta=0$}
The case of $\xi\neq0$ and $\Delta=0$ 
corresponds to $|B_{sy}|=|B_{sz}|\neq0$. 
In this case, 
\begin{equation}\label{pq_Delta0}
p=\sqrt{2\xi^2+|\xi|\sqrt{4\xi^2+a^2X^2}}, \ \ 
q=\frac{-\text{sgn}(\xi)aX}{2|\xi|+\sqrt{4\xi^2+a^2X^2}},
\end{equation}
and in particular, $|q|<1$. 
The double-angle formula of Jacobi elliptic function yields 
\begin{equation}\label{1+cd_nibaikaku}
\frac{\alpha}{X}=\frac{1}{2}[1+\text{cn}(pT,iq)\text{dn}(pT,iq)]=
\frac{\text{cn}^2(pT/2,iq)\text{dn}^2(pT/2,iq)
[1-q^2\text{sn}^4(pT/2,iq)]}
{[1+q^2\text{sn}^4(pT/2,iq)]^2}. 
\end{equation}
Because $|q|<1$, $\alpha=0$ holds if and only if 
$\text{cn}(pT/2,iq)=0$. Then, using 
\begin{equation}\label{Theta_xineq0Delta0}
\Theta=\mathscr{X}_1T
-q\xi\int_0^T
\frac{\text{sn}(p\tau,iq)^2}{1+\text{cn}(p\tau,iq)\text{dn}(p\tau,iq)}\text{d}\tau, 
\end{equation}
we obtain 
\begin{equation}\label{FGLM_xineq0Delta0}
\begin{split}
&f=-A\frac{\text{cn}(pT/2,iq)\text{dn}(pT/2,iq)
\sqrt{1-q^2\text{sn}^4(pT/2,iq)}}
{1+q^2\text{sn}^4(pT/2,iq)}\sin\Theta,\\
&g=A\frac{\text{cn}(pT/2,iq)\text{dn}(pT/2,iq)
\sqrt{1-q^2\text{sn}^4(pT/2,iq)}}
{1+q^2\text{sn}^4(pT/2,iq)}\cos\Theta,\\
&l=-\frac{2A\xi\text{sn}(pT/2,iq)}
{p\sqrt{1-q^2\text{sn}^4(pT/2,iq)}}
\left[
-\frac{aX}{2p}\text{sn}(pT,iq) \sin\Theta+\cos\Theta
\right],\\
&m=-\frac{2A\xi\text{sn}(pT/2,iq)}
{p\sqrt{1-q^2\text{sn}^4(pT/2,iq)}}
\left[
\frac{aX}{2p}\text{sn}(pT,iq) \cos\Theta+\sin\Theta
\right].\\
\end{split}
\end{equation}

\subsubsection{$f,g,l$, and $m$ for $\xi=0$}
This case corresponds to $B_{sy}=0$ or $B_{sz}=0$. 
We immediately see $\alpha=A^2,\beta=0,\gamma=0$, and 
obtain 
\begin{equation}\label{FGLM_xi0}
\begin{split}
&f=-A\sin\mathscr{X}_1T,\\
&g=A\cos\mathscr{X}_1T,\\
&l=0,\\
&m=0.\\
\end{split}
\end{equation}
The result contains the case of $\Delta=0$, {\it i.e.}, 
$B_s=0$.

\subsubsection{Approximation for $A\ll |B_{sy}|=|B_{sz}|$}
In the case of $A\ll |B_{sy}|=|B_{sz}|$, 
the solution given in 
Eqs. (A.19) and (A.20) can be approximated in a simple form. 
$p\approx 2|\xi|$ and $q\approx 0$ hold because $aX\ll |\xi|$. 
Therefore, 
\begin{equation}\label{}
\begin{split}
&f\approx-A\cos\xi T\sin\Theta,\\
&g\approx A\cos\xi T\cos\Theta,\\
&l\approx-A\sin\xi T\cos\Theta,\\
&m\approx-A\sin\xi T\sin\Theta,\\
\end{split}
\end{equation}
where 
\begin{equation}\label{}
\Theta\approx \left(\mathscr{X}_1+\frac{aX}{4}\right)T
-\frac{aX}{8\xi}\sin2\xi T
\approx \left(\mathscr{X}_1+\frac{aX}{4}\right)T. 
\end{equation}
The oscillating term can be discarded because 
its absolute value is much smaller than unity.

\subsection{Comparison to minimum corrective term}
We calculate for a short time scale. 
All of Eqs. (A.16), (A.20), 
and (A.21) express $f,g,l$, and $m$ give 
\begin{equation}\label{FGLM_small_T}
\begin{split}
&f=-A\mathscr{X}_1T+O(T^2),\\
&g=A+O(T^2),\\
&l=-A\xi T+O(T^2),\\
&m=O(T^2).\\
\end{split}
\end{equation}
Therefore, the main part of the corrective term 
in the short time scale are 
\begin{equation}\label{}
\begin{split}
&E_{ny}\approx-A\mathscr{X}_1\omega t\cos\omega t\sin kx,\\
&E_{nz}\approx-A\xi \omega t\cos\omega t\sin kx,\\
&B_{ny}\approx-A\xi \omega t\sin \omega t\cos kx,\\
&B_{nz}\approx A\mathscr{X}_1\omega t\sin \omega t\cos kx.\\
\end{split}
\end{equation}

\subsection{Calculation of the polarization}
For the calculation for Fig. 4 using realistic parameters, 
we derive the intensity ratio of the $y$ component 
to the entire electric field $I_y$ and the relative phase $\Psi_{y-z}$.

Each amplitude of 
the $y$ and $z$ component is given by 
$\sqrt{f^2+g^2}$ and $\sqrt{l^2+m^2}$, respectively, 
and the ratio $I_y$ is given by 
\begin{equation}\label{}
I_y=\frac{f^2+g^2}
{f^2+g^2+l^2+m^2}
=\frac{\alpha}{A^2}=\frac{1}{2}[1+\text{cn}(pT,iq)\text{dn}(pT,iq)]. 
\end{equation}

As for the relative phase, 
Eq. (A.20) for $f,g,l$, and $m$ and 
Eq. (A.19) for $\Theta$ yield  
\begin{equation}
\begin{split}
&E_y^{(\text{lp})}(x,t)=A
\sqrt{\frac{1+\text{cn}(pT,iq)\text{dn}(pT,iq)}{2}}
\sin(T-\Theta+\theta_c)
\sin kx,\\
&E_z^{(\text{lp})}(x,t)=A
\sqrt{\frac{1-\text{cn}(pT,iq)\text{dn}(pT,iq)}{2}}
\sin(T-\Theta-\Psi_0+\theta_s)
\sin kx,\\
\end{split}
\end{equation}
where the phase factors $\Psi_0 \in (0,\pi)$, 
$\theta_c$, and $\theta_s$ are defined as 
\begin{equation}\label{}
\begin{split}
&\sin\Psi_0=\frac{1}{\sqrt{1+[aX\text{sn}(pT,iq)/(2p)]^2}}, \ \ 
\cos\Psi_0=-\frac{aX\text{sn}(pT,iq)/(2p)}
{\sqrt{1+[aX\text{sn}(pT,iq)/(2p)]^2}},\\
&\theta_c=
\begin{cases}
0 & \boldsymbol{(}\text{cn}(pT/2,iq)\ge 0\boldsymbol{)}\\
\pi & \boldsymbol{(}\text{cn}(pT/2,iq)< 0\boldsymbol{)}\\
\end{cases}, \\
&\theta_s=
\begin{cases}
0 & \boldsymbol{(}\text{sn}(pT/2,iq)\ge 0\boldsymbol{)}\\
\pi & \boldsymbol{(}\text{sn}(pT/2,iq)< 0\boldsymbol{)}\\
\end{cases}. \\
\end{split}
\end{equation}
The phases of both components $\Psi_y$ and $\Psi_z$ 
can be defined as $\Psi_y=-\Theta+\theta_c$ and 
$\Psi_z=-\Theta-\Psi_0+\theta_s$, respectively. 
Then, the relative phase can be defined by 
\begin{equation}\label{}
\Psi_{y-z}=\Psi_0-\theta_{cs}, 
\end{equation}
where 
\begin{equation}\label{}
\theta_{cs}=
\begin{cases}
0 & \boldsymbol{(}\text{sn}(pT/2,iq)\text{cn}(pT/2,iq)
\ge 0\boldsymbol{)}\\
\pi & \boldsymbol{(}\text{sn}(pT/2,iq)\text{cn}(pT/2,iq)
< 0\boldsymbol{)}\\
\end{cases}. 
\end{equation}
We have defined $\Psi_{y-z}$ and $\theta_{cs}$ as above 
so that $\Psi_{y-z}$ ranges in $-\pi < \Psi_{y-z} < \pi$. 
The sign of $\Psi_{y-z}$ changes before and after 
at a time when $\text{sn}(pT/2,iq)=0$ or 
$\text{cn}(pT/2,iq)=0$ holds.

\begin{acknowledgments}
The author thanks 
Dr. M. Nakai, Dr. R. Kodama, Dr. K. Mima, and 
Dr. M. Fujita for discussions on the nonlinear 
electromagnetic wave and its experimental application, 
Dr. M. Uemoto for the advice on numerical calculations,  
and 
Mr. A. Watanabe for checking numerical calculations. 
The author quite appreciates Dr. J. Gabayno for checking 
the logical consistency of the text. 
\end{acknowledgments}


\begin{thebibliography}{51}%
\makeatletter
\providecommand \@ifxundefined [1]{%
 \@ifx{#1\undefined}
}%
\providecommand \@ifnum [1]{%
 \ifnum #1\expandafter \@firstoftwo
 \else \expandafter \@secondoftwo
 \fi
}%
\providecommand \@ifx [1]{%
 \ifx #1\expandafter \@firstoftwo
 \else \expandafter \@secondoftwo
 \fi
}%
\providecommand \natexlab [1]{#1}%
\providecommand \enquote  [1]{``#1''}%
\providecommand \bibnamefont  [1]{#1}%
\providecommand \bibfnamefont [1]{#1}%
\providecommand \citenamefont [1]{#1}%
\providecommand \href@noop [0]{\@secondoftwo}%
\providecommand \href [0]{\begingroup \@sanitize@url \@href}%
\providecommand \@href[1]{\@@startlink{#1}\@@href}%
\providecommand \@@href[1]{\endgroup#1\@@endlink}%
\providecommand \@sanitize@url [0]{\catcode `\\12\catcode `\$12\catcode
  `\&12\catcode `\#12\catcode `\^12\catcode `\_12\catcode `\%12\relax}%
\providecommand \@@startlink[1]{}%
\providecommand \@@endlink[0]{}%
\providecommand \url  [0]{\begingroup\@sanitize@url \@url }%
\providecommand \@url [1]{\endgroup\@href {#1}{\urlprefix }}%
\providecommand \urlprefix  [0]{URL }%
\providecommand \Eprint [0]{\href }%
\providecommand \doibase [0]{https://doi.org/}%
\providecommand \selectlanguage [0]{\@gobble}%
\providecommand \bibinfo  [0]{\@secondoftwo}%
\providecommand \bibfield  [0]{\@secondoftwo}%
\providecommand \translation [1]{[#1]}%
\providecommand \BibitemOpen [0]{}%
\providecommand \bibitemStop [0]{}%
\providecommand \bibitemNoStop [0]{.\EOS\space}%
\providecommand \EOS [0]{\spacefactor3000\relax}%
\providecommand \BibitemShut  [1]{\csname bibitem#1\endcsname}%
\let\auto@bib@innerbib\@empty
\bibitem [{\citenamefont {Heisenberg}\ and\ \citenamefont
  {Euler}(1936)}]{Heisenberg1936}%
  \BibitemOpen
  \bibfield  {author} {\bibinfo {author} {\bibfnamefont {W.}~\bibnamefont
  {Heisenberg}}\ and\ \bibinfo {author} {\bibfnamefont {H.}~\bibnamefont
  {Euler}},\ }\href {https://doi.org/10.1007/BF01343663} {\bibfield  {journal}
  {\bibinfo  {journal} {Zeitschrift f{\"u}r Physik}\ }\textbf {\bibinfo
  {volume} {98}},\ \bibinfo {pages} {714} (\bibinfo {year} {1936})}\BibitemShut
  {NoStop}%
\bibitem [{\citenamefont {Born}\ \emph {et~al.}(1934)\citenamefont {Born},
  \citenamefont {Infeld},\ and\ \citenamefont
  {Fowler}}]{doi:10.1098-rspa.1934.0059}%
  \BibitemOpen
  \bibfield  {author} {\bibinfo {author} {\bibfnamefont {M.}~\bibnamefont
  {Born}}, \bibinfo {author} {\bibfnamefont {L.}~\bibnamefont {Infeld}},\ and\
  \bibinfo {author} {\bibfnamefont {R.~H.}\ \bibnamefont {Fowler}},\ }\href
  {https://doi.org/10.1098/rspa.1934.0059} {\bibfield  {journal} {\bibinfo
  {journal} {Proceedings of the Royal Society of London. Series A, Containing
  Papers of a Mathematical and Physical Character}\ }\textbf {\bibinfo {volume}
  {144}},\ \bibinfo {pages} {425} (\bibinfo {year} {1934})},\ \Eprint
  {https://arxiv.org/abs/https://royalsocietypublishing.org/doi/pdf/10.1098/rspa.1934.0059}
  {https://royalsocietypublishing.org/doi/pdf/10.1098/rspa.1934.0059}
  \BibitemShut {NoStop}%
\bibitem [{\citenamefont {Plebanski}(1970)}]{Plebanski1970}%
  \BibitemOpen
  \bibfield  {author} {\bibinfo {author} {\bibfnamefont {J.}~\bibnamefont
  {Plebanski}},\ }\href {https://www.osti.gov/servlets/purl/4071071} {\emph
  {\bibinfo {title} {Lectures on non-linear electrodynamics: an extended
  version of lectures given at the Niels Bohr Institute and NORDITA,
  Copenhagen, in October 1968}}}\ (\bibinfo  {publisher} {Copenhagen :
  NORDITA},\ \bibinfo {year} {1970})\BibitemShut {NoStop}%
\bibitem [{\citenamefont {Di~Piazza}\ \emph {et~al.}(2012)\citenamefont
  {Di~Piazza}, \citenamefont {M\"uller}, \citenamefont {Hatsagortsyan},\ and\
  \citenamefont {Keitel}}]{RevModPhys.84.1177}%
  \BibitemOpen
  \bibfield  {author} {\bibinfo {author} {\bibfnamefont {A.}~\bibnamefont
  {Di~Piazza}}, \bibinfo {author} {\bibfnamefont {C.}~\bibnamefont {M\"uller}},
  \bibinfo {author} {\bibfnamefont {K.~Z.}\ \bibnamefont {Hatsagortsyan}},\
  and\ \bibinfo {author} {\bibfnamefont {C.~H.}\ \bibnamefont {Keitel}},\
  }\href {https://doi.org/10.1103/RevModPhys.84.1177} {\bibfield  {journal}
  {\bibinfo  {journal} {Rev. Mod. Phys.}\ }\textbf {\bibinfo {volume} {84}},\
  \bibinfo {pages} {1177} (\bibinfo {year} {2012})}\BibitemShut {NoStop}%
\bibitem [{\citenamefont {King}\ and\ \citenamefont
  {Heinzl}(2016)}]{king_heinzl_2016}%
  \BibitemOpen
  \bibfield  {author} {\bibinfo {author} {\bibfnamefont {B.}~\bibnamefont
  {King}}\ and\ \bibinfo {author} {\bibfnamefont {T.}~\bibnamefont {Heinzl}},\
  }\href {https://doi.org/10.1017/hpl.2016.1} {\bibfield  {journal} {\bibinfo
  {journal} {High Power Laser Science and Engineering}\ }\textbf {\bibinfo
  {volume} {4}},\ \bibinfo {pages} {e5} (\bibinfo {year} {2016})}\BibitemShut
  {NoStop}%
\bibitem [{\citenamefont {Heyl}\ and\ \citenamefont
  {Hernquist}(2005)}]{Heyl_2005}%
  \BibitemOpen
  \bibfield  {author} {\bibinfo {author} {\bibfnamefont {J.~S.}\ \bibnamefont
  {Heyl}}\ and\ \bibinfo {author} {\bibfnamefont {L.}~\bibnamefont
  {Hernquist}},\ }\href {https://doi.org/10.1086/425974} {\bibfield  {journal}
  {\bibinfo  {journal} {The Astrophysical Journal}\ }\textbf {\bibinfo {volume}
  {618}},\ \bibinfo {pages} {463} (\bibinfo {year} {2005})}\BibitemShut
  {NoStop}%
\bibitem [{\citenamefont {Shakeri}\ \emph {et~al.}(2017)\citenamefont
  {Shakeri}, \citenamefont {Haghighat},\ and\ \citenamefont
  {Xue}}]{Shakeri_2017}%
  \BibitemOpen
  \bibfield  {author} {\bibinfo {author} {\bibfnamefont {S.}~\bibnamefont
  {Shakeri}}, \bibinfo {author} {\bibfnamefont {M.}~\bibnamefont {Haghighat}},\
  and\ \bibinfo {author} {\bibfnamefont {S.-S.}\ \bibnamefont {Xue}},\ }\href
  {https://doi.org/10.1088/1475-7516/2017/10/014} {\bibfield  {journal}
  {\bibinfo  {journal} {Journal of Cosmology and Astroparticle Physics}\
  }\textbf {\bibinfo {volume} {2017}}\bibinfo  {number} { (10)},\ \bibinfo
  {pages} {014}}\BibitemShut {NoStop}%
\bibitem [{\citenamefont {Mignani}\ \emph {et~al.}(2016)\citenamefont
  {Mignani}, \citenamefont {Testa}, \citenamefont {Wu}, \citenamefont {Zane},
  \citenamefont {Gonzalez~Caniulef}, \citenamefont {Turolla},\ and\
  \citenamefont {Taverna}}]{10.1093/mnras/stw2798}%
  \BibitemOpen
\bibfield  {number} {  }\bibfield  {author} {\bibinfo {author} {\bibfnamefont
  {R.~P.}\ \bibnamefont {Mignani}}, \bibinfo {author} {\bibfnamefont
  {V.}~\bibnamefont {Testa}}, \bibinfo {author} {\bibfnamefont
  {K.}~\bibnamefont {Wu}}, \bibinfo {author} {\bibfnamefont {S.}~\bibnamefont
  {Zane}}, \bibinfo {author} {\bibfnamefont {D.}~\bibnamefont
  {Gonzalez~Caniulef}}, \bibinfo {author} {\bibfnamefont {R.}~\bibnamefont
  {Turolla}},\ and\ \bibinfo {author} {\bibfnamefont {R.}~\bibnamefont
  {Taverna}},\ }\href {https://doi.org/10.1093/mnras/stw2798} {\bibfield
  {journal} {\bibinfo  {journal} {Monthly Notices of the Royal Astronomical
  Society}\ }\textbf {\bibinfo {volume} {465}},\ \bibinfo {pages} {492}
  (\bibinfo {year} {2016})},\ \Eprint
  {https://arxiv.org/abs/http://oup.prod.sis.lan/mnras/article-pdf/465/1/492/8593962/stw2798.pdf}
  {http://oup.prod.sis.lan/mnras/article-pdf/465/1/492/8593962/stw2798.pdf}
  \BibitemShut {NoStop}%
\bibitem [{\citenamefont {Wichmann}\ and\ \citenamefont
  {Kroll}(1956)}]{PhysRev.101.843}%
  \BibitemOpen
  \bibfield  {author} {\bibinfo {author} {\bibfnamefont {E.~H.}\ \bibnamefont
  {Wichmann}}\ and\ \bibinfo {author} {\bibfnamefont {N.~M.}\ \bibnamefont
  {Kroll}},\ }\href {https://doi.org/10.1103/PhysRev.101.843} {\bibfield
  {journal} {\bibinfo  {journal} {Phys. Rev.}\ }\textbf {\bibinfo {volume}
  {101}},\ \bibinfo {pages} {843} (\bibinfo {year} {1956})}\BibitemShut
  {NoStop}%
\bibitem [{\citenamefont {Drebot}\ \emph {et~al.}(2017)\citenamefont {Drebot},
  \citenamefont {Micieli}, \citenamefont {Milotti}, \citenamefont {Petrillo},
  \citenamefont {Tassi},\ and\ \citenamefont
  {Serafini}}]{PhysRevAccelBeams.20.043402}%
  \BibitemOpen
  \bibfield  {author} {\bibinfo {author} {\bibfnamefont {I.}~\bibnamefont
  {Drebot}}, \bibinfo {author} {\bibfnamefont {D.}~\bibnamefont {Micieli}},
  \bibinfo {author} {\bibfnamefont {E.}~\bibnamefont {Milotti}}, \bibinfo
  {author} {\bibfnamefont {V.}~\bibnamefont {Petrillo}}, \bibinfo {author}
  {\bibfnamefont {E.}~\bibnamefont {Tassi}},\ and\ \bibinfo {author}
  {\bibfnamefont {L.}~\bibnamefont {Serafini}},\ }\href
  {https://doi.org/10.1103/PhysRevAccelBeams.20.043402} {\bibfield  {journal}
  {\bibinfo  {journal} {Phys. Rev. Accel. Beams}\ }\textbf {\bibinfo {volume}
  {20}},\ \bibinfo {pages} {043402} (\bibinfo {year} {2017})}\BibitemShut
  {NoStop}%
\bibitem [{\citenamefont {Uehling}(1935)}]{PhysRev.48.55}%
  \BibitemOpen
  \bibfield  {author} {\bibinfo {author} {\bibfnamefont {E.~A.}\ \bibnamefont
  {Uehling}},\ }\href {https://doi.org/10.1103/PhysRev.48.55} {\bibfield
  {journal} {\bibinfo  {journal} {Phys. Rev.}\ }\textbf {\bibinfo {volume}
  {48}},\ \bibinfo {pages} {55} (\bibinfo {year} {1935})}\BibitemShut {NoStop}%
\bibitem [{\citenamefont {Frolov}\ and\ \citenamefont
  {Wardlaw}(2012)}]{Frolov2012}%
  \BibitemOpen
  \bibfield  {author} {\bibinfo {author} {\bibfnamefont {A.~M.}\ \bibnamefont
  {Frolov}}\ and\ \bibinfo {author} {\bibfnamefont {D.~M.}\ \bibnamefont
  {Wardlaw}},\ }\href {https://doi.org/10.1140/epjb/e2012-30408-4} {\bibfield
  {journal} {\bibinfo  {journal} {The European Physical Journal B}\ }\textbf
  {\bibinfo {volume} {85}},\ \bibinfo {pages} {348} (\bibinfo {year}
  {2012})}\BibitemShut {NoStop}%
\bibitem [{\citenamefont {Frolov}\ and\ \citenamefont
  {Wardlaw}(2014)}]{FROLOV2014499}%
  \BibitemOpen
  \bibfield  {author} {\bibinfo {author} {\bibfnamefont {A.~M.}\ \bibnamefont
  {Frolov}}\ and\ \bibinfo {author} {\bibfnamefont {D.~M.}\ \bibnamefont
  {Wardlaw}},\ }\href
  {https://doi.org/https://doi.org/10.1016/j.jocs.2013.03.005} {\bibfield
  {journal} {\bibinfo  {journal} {Journal of Computational Science}\ }\textbf
  {\bibinfo {volume} {5}},\ \bibinfo {pages} {499 } (\bibinfo {year}
  {2014})}\BibitemShut {NoStop}%
\bibitem [{\citenamefont {Akmansoy}\ and\ \citenamefont
  {Medeiros}(2018)}]{Akmansoy2018}%
  \BibitemOpen
  \bibfield  {author} {\bibinfo {author} {\bibfnamefont {P.~N.}\ \bibnamefont
  {Akmansoy}}\ and\ \bibinfo {author} {\bibfnamefont {L.~G.}\ \bibnamefont
  {Medeiros}},\ }\href {https://doi.org/10.1140/epjc/s10052-018-5643-1}
  {\bibfield  {journal} {\bibinfo  {journal} {The European Physical Journal C}\
  }\textbf {\bibinfo {volume} {78}},\ \bibinfo {pages} {143} (\bibinfo {year}
  {2018})}\BibitemShut {NoStop}%
\bibitem [{\citenamefont {Carley}\ and\ \citenamefont
  {Kiessling}(2006)}]{PhysRevLett.96.030402}%
  \BibitemOpen
  \bibfield  {author} {\bibinfo {author} {\bibfnamefont {H.}~\bibnamefont
  {Carley}}\ and\ \bibinfo {author} {\bibfnamefont {M.~K.-H.}\ \bibnamefont
  {Kiessling}},\ }\href {https://doi.org/10.1103/PhysRevLett.96.030402}
  {\bibfield  {journal} {\bibinfo  {journal} {Phys. Rev. Lett.}\ }\textbf
  {\bibinfo {volume} {96}},\ \bibinfo {pages} {030402} (\bibinfo {year}
  {2006})}\BibitemShut {NoStop}%
\bibitem [{\citenamefont {Mazharimousavi}\ and\ \citenamefont
  {Halilsoy}(2012)}]{Mazharimousavi2012}%
  \BibitemOpen
  \bibfield  {author} {\bibinfo {author} {\bibfnamefont {S.~H.}\ \bibnamefont
  {Mazharimousavi}}\ and\ \bibinfo {author} {\bibfnamefont {M.}~\bibnamefont
  {Halilsoy}},\ }\href {https://doi.org/10.1007/s10701-011-9623-7} {\bibfield
  {journal} {\bibinfo  {journal} {Foundations of Physics}\ }\textbf {\bibinfo
  {volume} {42}},\ \bibinfo {pages} {524} (\bibinfo {year} {2012})}\BibitemShut
  {NoStop}%
\bibitem [{\citenamefont {Denisov}\ \emph {et~al.}(2006)\citenamefont
  {Denisov}, \citenamefont {Kravtsov},\ and\ \citenamefont
  {Krivchenkov}}]{Denisov2006}%
  \BibitemOpen
  \bibfield  {author} {\bibinfo {author} {\bibfnamefont {V.~I.}\ \bibnamefont
  {Denisov}}, \bibinfo {author} {\bibfnamefont {N.~V.}\ \bibnamefont
  {Kravtsov}},\ and\ \bibinfo {author} {\bibfnamefont {I.~V.}\ \bibnamefont
  {Krivchenkov}},\ }\href {https://doi.org/10.1134/S0030400X06050018}
  {\bibfield  {journal} {\bibinfo  {journal} {Optics and Spectroscopy}\
  }\textbf {\bibinfo {volume} {100}},\ \bibinfo {pages} {641} (\bibinfo {year}
  {2006})}\BibitemShut {NoStop}%
\bibitem [{\citenamefont {Ayon-Beato}\ and\ \citenamefont
  {Garc\'{i}a}(1999)}]{AYONBEATO199925}%
  \BibitemOpen
  \bibfield  {author} {\bibinfo {author} {\bibfnamefont {E.}~\bibnamefont
  {Ayon-Beato}}\ and\ \bibinfo {author} {\bibfnamefont {A.}~\bibnamefont
  {Garc\'{i}a}},\ }\href
  {https://doi.org/https://doi.org/10.1016/S0370-2693(99)01038-2} {\bibfield
  {journal} {\bibinfo  {journal} {Physics Letters B}\ }\textbf {\bibinfo
  {volume} {464}},\ \bibinfo {pages} {25} (\bibinfo {year} {1999})}\BibitemShut
  {NoStop}%
\bibitem [{\citenamefont {Bronnikov}(2001)}]{PhysRevD.63.044005}%
  \BibitemOpen
  \bibfield  {author} {\bibinfo {author} {\bibfnamefont {K.~A.}\ \bibnamefont
  {Bronnikov}},\ }\href {https://doi.org/10.1103/PhysRevD.63.044005} {\bibfield
   {journal} {\bibinfo  {journal} {Phys. Rev. D}\ }\textbf {\bibinfo {volume}
  {63}},\ \bibinfo {pages} {044005} (\bibinfo {year} {2001})}\BibitemShut
  {NoStop}%
\bibitem [{\citenamefont {Rizzo}\ \emph {et~al.}(2010)\citenamefont {Rizzo},
  \citenamefont {Dupays}, \citenamefont {Battesti}, \citenamefont
  {Fouch{\'{e}}},\ and\ \citenamefont {Rikken}}]{Rizzo_2010}%
  \BibitemOpen
  \bibfield  {author} {\bibinfo {author} {\bibfnamefont {C.}~\bibnamefont
  {Rizzo}}, \bibinfo {author} {\bibfnamefont {A.}~\bibnamefont {Dupays}},
  \bibinfo {author} {\bibfnamefont {R.}~\bibnamefont {Battesti}}, \bibinfo
  {author} {\bibfnamefont {M.}~\bibnamefont {Fouch{\'{e}}}},\ and\ \bibinfo
  {author} {\bibfnamefont {G.~L. J.~A.}\ \bibnamefont {Rikken}},\ }\href
  {https://doi.org/10.1209/0295-5075/90/64003} {\bibfield  {journal} {\bibinfo
  {journal} {{EPL} (Europhysics Letters)}\ }\textbf {\bibinfo {volume} {90}},\
  \bibinfo {pages} {64003} (\bibinfo {year} {2010})}\BibitemShut {NoStop}%
\bibitem [{\citenamefont {Lundstr\"om}\ \emph {et~al.}(2006)\citenamefont
  {Lundstr\"om}, \citenamefont {Brodin}, \citenamefont {Lundin}, \citenamefont
  {Marklund}, \citenamefont {Bingham}, \citenamefont {Collier}, \citenamefont
  {Mendon\ifmmode~\mbox{\c{c}}\else \c{c}\fi{}a},\ and\ \citenamefont
  {Norreys}}]{PhysRevLett.96.083602}%
  \BibitemOpen
  \bibfield  {author} {\bibinfo {author} {\bibfnamefont {E.}~\bibnamefont
  {Lundstr\"om}}, \bibinfo {author} {\bibfnamefont {G.}~\bibnamefont {Brodin}},
  \bibinfo {author} {\bibfnamefont {J.}~\bibnamefont {Lundin}}, \bibinfo
  {author} {\bibfnamefont {M.}~\bibnamefont {Marklund}}, \bibinfo {author}
  {\bibfnamefont {R.}~\bibnamefont {Bingham}}, \bibinfo {author} {\bibfnamefont
  {J.}~\bibnamefont {Collier}}, \bibinfo {author} {\bibfnamefont {J.~T.}\
  \bibnamefont {Mendon\ifmmode~\mbox{\c{c}}\else \c{c}\fi{}a}},\ and\ \bibinfo
  {author} {\bibfnamefont {P.}~\bibnamefont {Norreys}},\ }\href
  {https://doi.org/10.1103/PhysRevLett.96.083602} {\bibfield  {journal}
  {\bibinfo  {journal} {Phys. Rev. Lett.}\ }\textbf {\bibinfo {volume} {96}},\
  \bibinfo {pages} {083602} (\bibinfo {year} {2006})}\BibitemShut {NoStop}%
\bibitem [{\citenamefont {Lundin}\ \emph {et~al.}(2006)\citenamefont {Lundin},
  \citenamefont {Marklund}, \citenamefont {Lundstr\"om}, \citenamefont
  {Brodin}, \citenamefont {Collier}, \citenamefont {Bingham}, \citenamefont
  {Mendon\ifmmode~\mbox{\c{c}}\else \c{c}\fi{}a},\ and\ \citenamefont
  {Norreys}}]{PhysRevA.74.043821}%
  \BibitemOpen
  \bibfield  {author} {\bibinfo {author} {\bibfnamefont {J.}~\bibnamefont
  {Lundin}}, \bibinfo {author} {\bibfnamefont {M.}~\bibnamefont {Marklund}},
  \bibinfo {author} {\bibfnamefont {E.}~\bibnamefont {Lundstr\"om}}, \bibinfo
  {author} {\bibfnamefont {G.}~\bibnamefont {Brodin}}, \bibinfo {author}
  {\bibfnamefont {J.}~\bibnamefont {Collier}}, \bibinfo {author} {\bibfnamefont
  {R.}~\bibnamefont {Bingham}}, \bibinfo {author} {\bibfnamefont {J.~T.}\
  \bibnamefont {Mendon\ifmmode~\mbox{\c{c}}\else \c{c}\fi{}a}},\ and\ \bibinfo
  {author} {\bibfnamefont {P.}~\bibnamefont {Norreys}},\ }\href
  {https://doi.org/10.1103/PhysRevA.74.043821} {\bibfield  {journal} {\bibinfo
  {journal} {Phys. Rev. A}\ }\textbf {\bibinfo {volume} {74}},\ \bibinfo
  {pages} {043821} (\bibinfo {year} {2006})}\BibitemShut {NoStop}%
\bibitem [{\citenamefont {Sarazin}\ \emph {et~al.}(2016)\citenamefont
  {Sarazin}, \citenamefont {Couchot}, \citenamefont {Djannati-Ata{\"i}},
  \citenamefont {Guilbaud}, \citenamefont {Kazamias}, \citenamefont {Pittman},\
  and\ \citenamefont {Urban}}]{Sarazin2016}%
  \BibitemOpen
  \bibfield  {author} {\bibinfo {author} {\bibfnamefont {X.}~\bibnamefont
  {Sarazin}}, \bibinfo {author} {\bibfnamefont {F.}~\bibnamefont {Couchot}},
  \bibinfo {author} {\bibfnamefont {A.}~\bibnamefont {Djannati-Ata{\"i}}},
  \bibinfo {author} {\bibfnamefont {O.}~\bibnamefont {Guilbaud}}, \bibinfo
  {author} {\bibfnamefont {S.}~\bibnamefont {Kazamias}}, \bibinfo {author}
  {\bibfnamefont {M.}~\bibnamefont {Pittman}},\ and\ \bibinfo {author}
  {\bibfnamefont {M.}~\bibnamefont {Urban}},\ }\href
  {https://doi.org/10.1140/epjd/e2015-60428-5} {\bibfield  {journal} {\bibinfo
  {journal} {The European Physical Journal D}\ }\textbf {\bibinfo {volume}
  {70}},\ \bibinfo {pages} {13} (\bibinfo {year} {2016})}\BibitemShut {NoStop}%
\bibitem [{\citenamefont {Pinto Da~Souza}\ \emph {et~al.}(2006)\citenamefont
  {Pinto Da~Souza}, \citenamefont {Battesti}, \citenamefont {Robilliard},\ and\
  \citenamefont {Rizzo}}]{PintoDaSouza2006}%
  \BibitemOpen
  \bibfield  {author} {\bibinfo {author} {\bibfnamefont {B.}~\bibnamefont
  {Pinto Da~Souza}}, \bibinfo {author} {\bibfnamefont {R.}~\bibnamefont
  {Battesti}}, \bibinfo {author} {\bibfnamefont {C.}~\bibnamefont
  {Robilliard}},\ and\ \bibinfo {author} {\bibfnamefont {C.}~\bibnamefont
  {Rizzo}},\ }\href {https://doi.org/10.1140/epjd/e2006-00173-4} {\bibfield
  {journal} {\bibinfo  {journal} {The European Physical Journal D - Atomic,
  Molecular, Optical and Plasma Physics}\ }\textbf {\bibinfo {volume} {40}},\
  \bibinfo {pages} {445} (\bibinfo {year} {2006})}\BibitemShut {NoStop}%
\bibitem [{\citenamefont {Fouch\'e}\ \emph {et~al.}(2016)\citenamefont
  {Fouch\'e}, \citenamefont {Battesti},\ and\ \citenamefont
  {Rizzo}}]{PhysRevD.93.093020}%
  \BibitemOpen
  \bibfield  {author} {\bibinfo {author} {\bibfnamefont {M.}~\bibnamefont
  {Fouch\'e}}, \bibinfo {author} {\bibfnamefont {R.}~\bibnamefont {Battesti}},\
  and\ \bibinfo {author} {\bibfnamefont {C.}~\bibnamefont {Rizzo}},\ }\href
  {https://doi.org/10.1103/PhysRevD.93.093020} {\bibfield  {journal} {\bibinfo
  {journal} {Phys. Rev. D}\ }\textbf {\bibinfo {volume} {93}},\ \bibinfo
  {pages} {093020} (\bibinfo {year} {2016})}\BibitemShut {NoStop}%
\bibitem [{\citenamefont {Fouch\'e}\ \emph {et~al.}(2017)\citenamefont
  {Fouch\'e}, \citenamefont {Battesti},\ and\ \citenamefont
  {Rizzo}}]{PhysRevD.95.099902}%
  \BibitemOpen
  \bibfield  {author} {\bibinfo {author} {\bibfnamefont {M.}~\bibnamefont
  {Fouch\'e}}, \bibinfo {author} {\bibfnamefont {R.}~\bibnamefont {Battesti}},\
  and\ \bibinfo {author} {\bibfnamefont {C.}~\bibnamefont {Rizzo}},\ }\href
  {https://doi.org/10.1103/PhysRevD.95.099902} {\bibfield  {journal} {\bibinfo
  {journal} {Phys. Rev. D}\ }\textbf {\bibinfo {volume} {95}},\ \bibinfo
  {pages} {099902} (\bibinfo {year} {2017})}\BibitemShut {NoStop}%
\bibitem [{\citenamefont {Battesti}\ and\ \citenamefont
  {Rizzo}(2012)}]{Battesti_2012}%
  \BibitemOpen
  \bibfield  {author} {\bibinfo {author} {\bibfnamefont {R.}~\bibnamefont
  {Battesti}}\ and\ \bibinfo {author} {\bibfnamefont {C.}~\bibnamefont
  {Rizzo}},\ }\href {https://doi.org/10.1088/0034-4885/76/1/016401} {\bibfield
  {journal} {\bibinfo  {journal} {Reports on Progress in Physics}\ }\textbf
  {\bibinfo {volume} {76}},\ \bibinfo {pages} {016401} (\bibinfo {year}
  {2012})}\BibitemShut {NoStop}%
\bibitem [{\citenamefont {Bernard}\ \emph {et~al.}(2000)\citenamefont
  {Bernard}, \citenamefont {Moulin}, \citenamefont {Amiranoff}, \citenamefont
  {Braun}, \citenamefont {Chambaret}, \citenamefont {Darpentigny},
  \citenamefont {Grillon}, \citenamefont {Ranc},\ and\ \citenamefont
  {Perrone}}]{Bernard2000}%
  \BibitemOpen
  \bibfield  {author} {\bibinfo {author} {\bibfnamefont {D.}~\bibnamefont
  {Bernard}}, \bibinfo {author} {\bibfnamefont {F.}~\bibnamefont {Moulin}},
  \bibinfo {author} {\bibfnamefont {F.}~\bibnamefont {Amiranoff}}, \bibinfo
  {author} {\bibfnamefont {A.}~\bibnamefont {Braun}}, \bibinfo {author}
  {\bibfnamefont {J.}~\bibnamefont {Chambaret}}, \bibinfo {author}
  {\bibfnamefont {G.}~\bibnamefont {Darpentigny}}, \bibinfo {author}
  {\bibfnamefont {G.}~\bibnamefont {Grillon}}, \bibinfo {author} {\bibfnamefont
  {S.}~\bibnamefont {Ranc}},\ and\ \bibinfo {author} {\bibfnamefont
  {F.}~\bibnamefont {Perrone}},\ }\href {https://doi.org/10.1007/s100530050535}
  {\bibfield  {journal} {\bibinfo  {journal} {The European Physical Journal D -
  Atomic, Molecular, Optical and Plasma Physics}\ }\textbf {\bibinfo {volume}
  {10}},\ \bibinfo {pages} {141} (\bibinfo {year} {2000})}\BibitemShut
  {NoStop}%
\bibitem [{\citenamefont {Della~Valle}\ \emph {et~al.}(2016)\citenamefont
  {Della~Valle}, \citenamefont {Ejlli}, \citenamefont {Gastaldi}, \citenamefont
  {Messineo}, \citenamefont {Milotti}, \citenamefont {Pengo}, \citenamefont
  {Ruoso},\ and\ \citenamefont {Zavattini}}]{DellaValle2016}%
  \BibitemOpen
  \bibfield  {author} {\bibinfo {author} {\bibfnamefont {F.}~\bibnamefont
  {Della~Valle}}, \bibinfo {author} {\bibfnamefont {A.}~\bibnamefont {Ejlli}},
  \bibinfo {author} {\bibfnamefont {U.}~\bibnamefont {Gastaldi}}, \bibinfo
  {author} {\bibfnamefont {G.}~\bibnamefont {Messineo}}, \bibinfo {author}
  {\bibfnamefont {E.}~\bibnamefont {Milotti}}, \bibinfo {author} {\bibfnamefont
  {R.}~\bibnamefont {Pengo}}, \bibinfo {author} {\bibfnamefont
  {G.}~\bibnamefont {Ruoso}},\ and\ \bibinfo {author} {\bibfnamefont
  {G.}~\bibnamefont {Zavattini}},\ }\href
  {https://doi.org/10.1140/epjc/s10052-015-3869-8} {\bibfield  {journal}
  {\bibinfo  {journal} {The European Physical Journal C}\ }\textbf {\bibinfo
  {volume} {76}},\ \bibinfo {pages} {24} (\bibinfo {year} {2016})}\BibitemShut
  {NoStop}%
\bibitem [{\citenamefont {Fan}\ \emph {et~al.}(2017)\citenamefont {Fan},
  \citenamefont {Kamioka}, \citenamefont {Inada}, \citenamefont {Yamazaki},
  \citenamefont {Namba}, \citenamefont {Asai}, \citenamefont {Omachi},
  \citenamefont {Yoshioka}, \citenamefont {Kuwata-Gonokami}, \citenamefont
  {Matsuo}, \citenamefont {Kawaguchi}, \citenamefont {Kindo},\ and\
  \citenamefont {Nojiri}}]{Fan2017}%
  \BibitemOpen
  \bibfield  {author} {\bibinfo {author} {\bibfnamefont {X.}~\bibnamefont
  {Fan}}, \bibinfo {author} {\bibfnamefont {S.}~\bibnamefont {Kamioka}},
  \bibinfo {author} {\bibfnamefont {T.}~\bibnamefont {Inada}}, \bibinfo
  {author} {\bibfnamefont {T.}~\bibnamefont {Yamazaki}}, \bibinfo {author}
  {\bibfnamefont {T.}~\bibnamefont {Namba}}, \bibinfo {author} {\bibfnamefont
  {S.}~\bibnamefont {Asai}}, \bibinfo {author} {\bibfnamefont {J.}~\bibnamefont
  {Omachi}}, \bibinfo {author} {\bibfnamefont {K.}~\bibnamefont {Yoshioka}},
  \bibinfo {author} {\bibfnamefont {M.}~\bibnamefont {Kuwata-Gonokami}},
  \bibinfo {author} {\bibfnamefont {A.}~\bibnamefont {Matsuo}}, \bibinfo
  {author} {\bibfnamefont {K.}~\bibnamefont {Kawaguchi}}, \bibinfo {author}
  {\bibfnamefont {K.}~\bibnamefont {Kindo}},\ and\ \bibinfo {author}
  {\bibfnamefont {H.}~\bibnamefont {Nojiri}},\ }\href
  {https://doi.org/10.1140/epjd/e2017-80290-7} {\bibfield  {journal} {\bibinfo
  {journal} {The European Physical Journal D}\ }\textbf {\bibinfo {volume}
  {71}},\ \bibinfo {pages} {308} (\bibinfo {year} {2017})}\BibitemShut
  {NoStop}%
\bibitem [{\citenamefont {Cad{\`e}ne}\ \emph {et~al.}(2014)\citenamefont
  {Cad{\`e}ne}, \citenamefont {Berceau}, \citenamefont {Fouch{\'e}},
  \citenamefont {Battesti},\ and\ \citenamefont {Rizzo}}]{Cadene2014}%
  \BibitemOpen
  \bibfield  {author} {\bibinfo {author} {\bibfnamefont {A.}~\bibnamefont
  {Cad{\`e}ne}}, \bibinfo {author} {\bibfnamefont {P.}~\bibnamefont {Berceau}},
  \bibinfo {author} {\bibfnamefont {M.}~\bibnamefont {Fouch{\'e}}}, \bibinfo
  {author} {\bibfnamefont {R.}~\bibnamefont {Battesti}},\ and\ \bibinfo
  {author} {\bibfnamefont {C.}~\bibnamefont {Rizzo}},\ }\href
  {https://doi.org/10.1140/epjd/e2013-40725-9} {\bibfield  {journal} {\bibinfo
  {journal} {The European Physical Journal D}\ }\textbf {\bibinfo {volume}
  {68}},\ \bibinfo {pages} {16} (\bibinfo {year} {2014})}\BibitemShut {NoStop}%
\bibitem [{\citenamefont {Karbstein}(2020)}]{particles3010005}%
  \BibitemOpen
  \bibfield  {author} {\bibinfo {author} {\bibfnamefont {F.}~\bibnamefont
  {Karbstein}},\ }\href {https://doi.org/10.3390/particles3010005} {\bibfield
  {journal} {\bibinfo  {journal} {Particles}\ }\textbf {\bibinfo {volume}
  {3}},\ \bibinfo {pages} {39} (\bibinfo {year} {2020})}\BibitemShut {NoStop}%
\bibitem [{\citenamefont {Marklund}\ and\ \citenamefont
  {Shukla}(2006)}]{RevModPhys.78.591}%
  \BibitemOpen
  \bibfield  {author} {\bibinfo {author} {\bibfnamefont {M.}~\bibnamefont
  {Marklund}}\ and\ \bibinfo {author} {\bibfnamefont {P.~K.}\ \bibnamefont
  {Shukla}},\ }\href {https://doi.org/10.1103/RevModPhys.78.591} {\bibfield
  {journal} {\bibinfo  {journal} {Rev. Mod. Phys.}\ }\textbf {\bibinfo {volume}
  {78}},\ \bibinfo {pages} {591} (\bibinfo {year} {2006})}\BibitemShut
  {NoStop}%
\bibitem [{\citenamefont {Rozanov}(1998)}]{Rozanov1998}%
  \BibitemOpen
  \bibfield  {author} {\bibinfo {author} {\bibfnamefont {N.~N.}\ \bibnamefont
  {Rozanov}},\ }\href {https://doi.org/10.1134/1.558454} {\bibfield  {journal}
  {\bibinfo  {journal} {Journal of Experimental and Theoretical Physics}\
  }\textbf {\bibinfo {volume} {86}},\ \bibinfo {pages} {284} (\bibinfo {year}
  {1998})}\BibitemShut {NoStop}%
\bibitem [{\citenamefont {Uno}\ \emph {et~al.}(2016)\citenamefont {Uno},
  \citenamefont {Ka},\ and\ \citenamefont {Arima}}]{FDTD_Book}%
  \BibitemOpen
  \bibfield  {author} {\bibinfo {author} {\bibfnamefont {T.}~\bibnamefont
  {Uno}}, \bibinfo {author} {\bibfnamefont {I.}~\bibnamefont {Ka}},\ and\
  \bibinfo {author} {\bibfnamefont {T.}~\bibnamefont {Arima}},\ }\href@noop {}
  {\emph {\bibinfo {title} {FDTD Method for Computational Electromagnetics}}},\
  \bibinfo {edition} {1st}\ ed.\ (\bibinfo  {publisher} {CORONA PUBLISHING
  CO.,LTD.},\ \bibinfo {address} {Tokyo Japan},\ \bibinfo {year}
  {2016})\BibitemShut {NoStop}%
\bibitem [{\citenamefont {Shibata}(2020)}]{Shibata2020}%
  \BibitemOpen
  \bibfield  {author} {\bibinfo {author} {\bibfnamefont {K.}~\bibnamefont
  {Shibata}},\ }\href {https://doi.org/10.1140/epjd/e2020-10420-1} {\bibfield
  {journal} {\bibinfo  {journal} {The European Physical Journal D}\ }\textbf
  {\bibinfo {volume} {74}},\ \bibinfo {pages} {215} (\bibinfo {year}
  {2020})}\BibitemShut {NoStop}%
\bibitem [{\citenamefont {Shibata}(2021)}]{Shibata2021EPJD}%
  \BibitemOpen
  \bibfield  {author} {\bibinfo {author} {\bibfnamefont {K.}~\bibnamefont
  {Shibata}},\ }\href {https://doi.org/10.1140/epjd/s10053-021-00181-w}
  {\bibfield  {journal} {\bibinfo  {journal} {The European Physical Journal D}\
  }\textbf {\bibinfo {volume} {75}},\ \bibinfo {pages} {169} (\bibinfo {year}
  {2021})}\BibitemShut {NoStop}%
\bibitem [{\citenamefont {Brodin}\ \emph {et~al.}(2001)\citenamefont {Brodin},
  \citenamefont {Marklund},\ and\ \citenamefont
  {Stenflo}}]{PhysRevLett.87.171801}%
  \BibitemOpen
  \bibfield  {author} {\bibinfo {author} {\bibfnamefont {G.}~\bibnamefont
  {Brodin}}, \bibinfo {author} {\bibfnamefont {M.}~\bibnamefont {Marklund}},\
  and\ \bibinfo {author} {\bibfnamefont {L.}~\bibnamefont {Stenflo}},\ }\href
  {https://doi.org/10.1103/PhysRevLett.87.171801} {\bibfield  {journal}
  {\bibinfo  {journal} {Phys. Rev. Lett.}\ }\textbf {\bibinfo {volume} {87}},\
  \bibinfo {pages} {171801} (\bibinfo {year} {2001})}\BibitemShut {NoStop}%
\bibitem [{\citenamefont {Schwinger}(1951)}]{PhysRev.82.664}%
  \BibitemOpen
  \bibfield  {author} {\bibinfo {author} {\bibfnamefont {J.}~\bibnamefont
  {Schwinger}},\ }\href {https://doi.org/10.1103/PhysRev.82.664} {\bibfield
  {journal} {\bibinfo  {journal} {Phys. Rev.}\ }\textbf {\bibinfo {volume}
  {82}},\ \bibinfo {pages} {664} (\bibinfo {year} {1951})}\BibitemShut
  {NoStop}%
\bibitem [{\citenamefont {Katori}\ \emph {et~al.}(2015)\citenamefont {Katori},
  \citenamefont {Ovsiannikov}, \citenamefont {Marmo},\ and\ \citenamefont
  {Palchikov}}]{PhysRevA.91.052503}%
  \BibitemOpen
  \bibfield  {author} {\bibinfo {author} {\bibfnamefont {H.}~\bibnamefont
  {Katori}}, \bibinfo {author} {\bibfnamefont {V.~D.}\ \bibnamefont
  {Ovsiannikov}}, \bibinfo {author} {\bibfnamefont {S.~I.}\ \bibnamefont
  {Marmo}},\ and\ \bibinfo {author} {\bibfnamefont {V.~G.}\ \bibnamefont
  {Palchikov}},\ }\href {https://doi.org/10.1103/PhysRevA.91.052503} {\bibfield
   {journal} {\bibinfo  {journal} {Phys. Rev. A}\ }\textbf {\bibinfo {volume}
  {91}},\ \bibinfo {pages} {052503} (\bibinfo {year} {2015})}\BibitemShut
  {NoStop}%
\bibitem [{\citenamefont {Durrell}\ \emph {et~al.}(2014)\citenamefont
  {Durrell}, \citenamefont {Dennis}, \citenamefont {Jaroszynski}, \citenamefont
  {Ainslie}, \citenamefont {Palmer}, \citenamefont {Shi}, \citenamefont
  {Campbell}, \citenamefont {Hull}, \citenamefont {Strasik}, \citenamefont
  {Hellstrom},\ and\ \citenamefont {Cardwell}}]{Durrell_2014}%
  \BibitemOpen
  \bibfield  {author} {\bibinfo {author} {\bibfnamefont {J.~H.}\ \bibnamefont
  {Durrell}}, \bibinfo {author} {\bibfnamefont {A.~R.}\ \bibnamefont {Dennis}},
  \bibinfo {author} {\bibfnamefont {J.}~\bibnamefont {Jaroszynski}}, \bibinfo
  {author} {\bibfnamefont {M.~D.}\ \bibnamefont {Ainslie}}, \bibinfo {author}
  {\bibfnamefont {K.~G.~B.}\ \bibnamefont {Palmer}}, \bibinfo {author}
  {\bibfnamefont {Y.-H.}\ \bibnamefont {Shi}}, \bibinfo {author} {\bibfnamefont
  {A.~M.}\ \bibnamefont {Campbell}}, \bibinfo {author} {\bibfnamefont
  {J.}~\bibnamefont {Hull}}, \bibinfo {author} {\bibfnamefont {M.}~\bibnamefont
  {Strasik}}, \bibinfo {author} {\bibfnamefont {E.~E.}\ \bibnamefont
  {Hellstrom}},\ and\ \bibinfo {author} {\bibfnamefont {D.~A.}\ \bibnamefont
  {Cardwell}},\ }\href {https://doi.org/10.1088/0953-2048/27/8/082001}
  {\bibfield  {journal} {\bibinfo  {journal} {Superconductor Science and
  Technology}\ }\textbf {\bibinfo {volume} {27}},\ \bibinfo {pages} {082001}
  (\bibinfo {year} {2014})}\BibitemShut {NoStop}%
\bibitem [{\citenamefont {Durrell}\ \emph {et~al.}(2018)\citenamefont
  {Durrell}, \citenamefont {Ainslie}, \citenamefont {Zhou}, \citenamefont
  {Vanderbemden}, \citenamefont {Bradshaw}, \citenamefont {Speller},
  \citenamefont {Filipenko},\ and\ \citenamefont {Cardwell}}]{Durrell_2018}%
  \BibitemOpen
  \bibfield  {author} {\bibinfo {author} {\bibfnamefont {J.~H.}\ \bibnamefont
  {Durrell}}, \bibinfo {author} {\bibfnamefont {M.~D.}\ \bibnamefont
  {Ainslie}}, \bibinfo {author} {\bibfnamefont {D.}~\bibnamefont {Zhou}},
  \bibinfo {author} {\bibfnamefont {P.}~\bibnamefont {Vanderbemden}}, \bibinfo
  {author} {\bibfnamefont {T.}~\bibnamefont {Bradshaw}}, \bibinfo {author}
  {\bibfnamefont {S.}~\bibnamefont {Speller}}, \bibinfo {author} {\bibfnamefont
  {M.}~\bibnamefont {Filipenko}},\ and\ \bibinfo {author} {\bibfnamefont
  {D.~A.}\ \bibnamefont {Cardwell}},\ }\href
  {https://doi.org/10.1088/1361-6668/aad7ce} {\bibfield  {journal} {\bibinfo
  {journal} {Superconductor Science and Technology}\ }\textbf {\bibinfo
  {volume} {31}},\ \bibinfo {pages} {103501} (\bibinfo {year}
  {2018})}\BibitemShut {NoStop}%
\bibitem [{\citenamefont {Mukherjee}\ \emph {et~al.}(2019)\citenamefont
  {Mukherjee}, \citenamefont {Ishida}, \citenamefont {Hagen},\ and\
  \citenamefont {Otani}}]{Mukherjee2019}%
  \BibitemOpen
  \bibfield  {author} {\bibinfo {author} {\bibfnamefont {P.}~\bibnamefont
  {Mukherjee}}, \bibinfo {author} {\bibfnamefont {S.}~\bibnamefont {Ishida}},
  \bibinfo {author} {\bibfnamefont {N.}~\bibnamefont {Hagen}},\ and\ \bibinfo
  {author} {\bibfnamefont {Y.}~\bibnamefont {Otani}},\ }\href
  {https://doi.org/10.1007/s10043-018-0475-7} {\bibfield  {journal} {\bibinfo
  {journal} {Optical Review}\ }\textbf {\bibinfo {volume} {26}},\ \bibinfo
  {pages} {23} (\bibinfo {year} {2019})}\BibitemShut {NoStop}%
\bibitem [{\citenamefont {Majkic}\ \emph {et~al.}(2020)\citenamefont {Majkic},
  \citenamefont {Pratap}, \citenamefont {Paidpilli}, \citenamefont {Galstyan},
  \citenamefont {Kochat}, \citenamefont {Goel}, \citenamefont {Kar},
  \citenamefont {Jaroszynski}, \citenamefont {Abraimov},\ and\ \citenamefont
  {Selvamanickam}}]{Majkic_2020}%
  \BibitemOpen
  \bibfield  {author} {\bibinfo {author} {\bibfnamefont {G.}~\bibnamefont
  {Majkic}}, \bibinfo {author} {\bibfnamefont {R.}~\bibnamefont {Pratap}},
  \bibinfo {author} {\bibfnamefont {M.}~\bibnamefont {Paidpilli}}, \bibinfo
  {author} {\bibfnamefont {E.}~\bibnamefont {Galstyan}}, \bibinfo {author}
  {\bibfnamefont {M.}~\bibnamefont {Kochat}}, \bibinfo {author} {\bibfnamefont
  {C.}~\bibnamefont {Goel}}, \bibinfo {author} {\bibfnamefont {S.}~\bibnamefont
  {Kar}}, \bibinfo {author} {\bibfnamefont {J.}~\bibnamefont {Jaroszynski}},
  \bibinfo {author} {\bibfnamefont {D.}~\bibnamefont {Abraimov}},\ and\
  \bibinfo {author} {\bibfnamefont {V.}~\bibnamefont {Selvamanickam}},\ }\href
  {https://doi.org/10.1088/1361-6668/ab9541} {\bibfield  {journal} {\bibinfo
  {journal} {Superconductor Science and Technology}\ }\textbf {\bibinfo
  {volume} {33}},\ \bibinfo {pages} {07LT03} (\bibinfo {year}
  {2020})}\BibitemShut {NoStop}%
\bibitem [{\citenamefont {Hirose}\ \emph {et~al.}(2020)\citenamefont {Hirose},
  \citenamefont {Billingsley}, \citenamefont {Zhang}, \citenamefont {Yamamoto},
  \citenamefont {Pinard}, \citenamefont {Michel}, \citenamefont {Forest},
  \citenamefont {Reichman},\ and\ \citenamefont
  {Gross}}]{PhysRevApplied.14.014021}%
  \BibitemOpen
  \bibfield  {author} {\bibinfo {author} {\bibfnamefont {E.}~\bibnamefont
  {Hirose}}, \bibinfo {author} {\bibfnamefont {G.}~\bibnamefont {Billingsley}},
  \bibinfo {author} {\bibfnamefont {L.}~\bibnamefont {Zhang}}, \bibinfo
  {author} {\bibfnamefont {H.}~\bibnamefont {Yamamoto}}, \bibinfo {author}
  {\bibfnamefont {L.}~\bibnamefont {Pinard}}, \bibinfo {author} {\bibfnamefont
  {C.}~\bibnamefont {Michel}}, \bibinfo {author} {\bibfnamefont
  {D.}~\bibnamefont {Forest}}, \bibinfo {author} {\bibfnamefont
  {B.}~\bibnamefont {Reichman}},\ and\ \bibinfo {author} {\bibfnamefont
  {M.}~\bibnamefont {Gross}},\ }\href
  {https://doi.org/10.1103/PhysRevApplied.14.014021} {\bibfield  {journal}
  {\bibinfo  {journal} {Phys. Rev. Applied}\ }\textbf {\bibinfo {volume}
  {14}},\ \bibinfo {pages} {014021} (\bibinfo {year} {2020})}\BibitemShut
  {NoStop}%
\bibitem [{\citenamefont {Akutsu}\ \emph {et~al.}(2020)\citenamefont {Akutsu},
  \citenamefont {Ando}, \citenamefont {Arai}, \citenamefont {Arai},
  \citenamefont {Araki}, \citenamefont {Araya}, \citenamefont {Aritomi},
  \citenamefont {Aso}, \citenamefont {Bae}, \citenamefont {Bae}, \citenamefont
  {Baiotti}, \citenamefont {Bajpai}, \citenamefont {Barton}, \citenamefont
  {Cannon}, \citenamefont {Capocasa}, \citenamefont {Chan}, \citenamefont
  {Chen}, \citenamefont {Chen}, \citenamefont {Chen}, \citenamefont {Chu},
  \citenamefont {Chu}, \citenamefont {Eguchi}, \citenamefont {Enomoto},
  \citenamefont {Flaminio}, \citenamefont {Fujii}, \citenamefont {Fukunaga},
  \citenamefont {Fukushima}, \citenamefont {Ge}, \citenamefont {Hagiwara},
  \citenamefont {Haino}, \citenamefont {Hasegawa}, \citenamefont {Hayakawa},
  \citenamefont {Hayama}, \citenamefont {Himemoto}, \citenamefont {Hiranuma},
  \citenamefont {Hirata}, \citenamefont {Hirose}, \citenamefont {Hong},
  \citenamefont {Hsieh}, \citenamefont {Huang}, \citenamefont {Huang},
  \citenamefont {Huang}, \citenamefont {Ikenoue}, \citenamefont {Imam},
  \citenamefont {Inayoshi}, \citenamefont {Inoue}, \citenamefont {Ioka},
  \citenamefont {Itoh}, \citenamefont {Izumi}, \citenamefont {Jung},
  \citenamefont {Jung}, \citenamefont {Kajita}, \citenamefont {Kamiizumi},
  \citenamefont {Kanda}, \citenamefont {Kang}, \citenamefont {Kawaguchi},
  \citenamefont {Kawai}, \citenamefont {Kawasaki}, \citenamefont {Kim},
  \citenamefont {Kim}, \citenamefont {Kim}, \citenamefont {Kim}, \citenamefont
  {Kimura}, \citenamefont {Kita}, \citenamefont {Kitazawa}, \citenamefont
  {Kojima}, \citenamefont {Kokeyama}, \citenamefont {Komori}, \citenamefont
  {Kong}, \citenamefont {Kotake}, \citenamefont {Kozakai}, \citenamefont
  {Kozu}, \citenamefont {Kumar}, \citenamefont {Kume}, \citenamefont {Kuo},
  \citenamefont {Kuo}, \citenamefont {Kuroyanagi}, \citenamefont {Kusayanagi},
  \citenamefont {Kwak}, \citenamefont {Lee}, \citenamefont {Lee}, \citenamefont
  {Lee}, \citenamefont {Leonardi}, \citenamefont {Lin}, \citenamefont {Lin},
  \citenamefont {Lin}, \citenamefont {Liu}, \citenamefont {Luo}, \citenamefont
  {Marchio}, \citenamefont {Michimura}, \citenamefont {Mio}, \citenamefont
  {Miyakawa}, \citenamefont {Miyamoto}, \citenamefont {Miyazaki}, \citenamefont
  {Miyo}, \citenamefont {Miyoki}, \citenamefont {Morisaki}, \citenamefont
  {Moriwaki}, \citenamefont {Nagano}, \citenamefont {Nagano}, \citenamefont
  {Nakamura}, \citenamefont {Nakano}, \citenamefont {Nakano}, \citenamefont
  {Nakashima}, \citenamefont {Narikawa}, \citenamefont {Negishi}, \citenamefont
  {Ni}, \citenamefont {Nishizawa}, \citenamefont {Obuchi}, \citenamefont
  {Ogaki}, \citenamefont {Oh}, \citenamefont {Oh}, \citenamefont {Ohashi},
  \citenamefont {Ohishi}, \citenamefont {Ohkawa}, \citenamefont {Okutomi},
  \citenamefont {Oohara}, \citenamefont {Ooi}, \citenamefont {Oshino},
  \citenamefont {Pan}, \citenamefont {Pang}, \citenamefont {Park},
  \citenamefont {Arellano}, \citenamefont {Pinto}, \citenamefont {Sago},
  \citenamefont {Saito}, \citenamefont {Saito}, \citenamefont {Sakai},
  \citenamefont {Sakai}, \citenamefont {Sakuno}, \citenamefont {Sato},
  \citenamefont {Sato}, \citenamefont {Sawada}, \citenamefont {Sekiguchi},
  \citenamefont {Sekiguchi}, \citenamefont {Shibagaki}, \citenamefont
  {Shimizu}, \citenamefont {Shimoda}, \citenamefont {Shimode}, \citenamefont
  {Shinkai}, \citenamefont {Shishido}, \citenamefont {Shoda}, \citenamefont
  {Somiya}, \citenamefont {Son}, \citenamefont {Sotani}, \citenamefont
  {Sugimoto}, \citenamefont {Suzuki}, \citenamefont {Suzuki}, \citenamefont
  {Tagoshi}, \citenamefont {Takahashi}, \citenamefont {Takahashi},
  \citenamefont {Takamori}, \citenamefont {Takano}, \citenamefont {Takeda},
  \citenamefont {Takeda}, \citenamefont {Tanaka}, \citenamefont {Tanaka},
  \citenamefont {Tanaka}, \citenamefont {Tanaka}, \citenamefont {Tanaka},
  \citenamefont {Tanioka}, \citenamefont {Tapia San~Martin}, \citenamefont
  {Telada}, \citenamefont {Tomaru}, \citenamefont {Tomigami}, \citenamefont
  {Tomura}, \citenamefont {Travasso}, \citenamefont {Trozzo}, \citenamefont
  {Tsang}, \citenamefont {Tsubono}, \citenamefont {Tsuchida}, \citenamefont
  {Tsuzuki}, \citenamefont {Tuyenbayev}, \citenamefont {Uchikata},
  \citenamefont {Uchiyama}, \citenamefont {Ueda}, \citenamefont {Uehara},
  \citenamefont {Ueno}, \citenamefont {Ueshima}, \citenamefont {Uraguchi},
  \citenamefont {Ushiba}, \citenamefont {van Putten}, \citenamefont {Vocca},
  \citenamefont {Wang}, \citenamefont {Wu}, \citenamefont {Wu}, \citenamefont
  {Wu}, \citenamefont {Xu}, \citenamefont {Yamada}, \citenamefont {Yamamoto},
  \citenamefont {Yamamoto}, \citenamefont {Yamamoto}, \citenamefont {Yokogawa},
  \citenamefont {Yokoyama}, \citenamefont {Yokozawa}, \citenamefont {Yoshioka},
  \citenamefont {Yuzurihara}, \citenamefont {Zeidler}, \citenamefont {Zhao},\
  and\ \citenamefont {Zhu}}]{10.1093/ptep/ptaa125}%
  \BibitemOpen
  \bibfield  {author} {\bibinfo {author} {\bibfnamefont {T.}~\bibnamefont
  {Akutsu}}, \bibinfo {author} {\bibfnamefont {M.}~\bibnamefont {Ando}},
  \bibinfo {author} {\bibfnamefont {K.}~\bibnamefont {Arai}}, \bibinfo {author}
  {\bibfnamefont {Y.}~\bibnamefont {Arai}}, \bibinfo {author} {\bibfnamefont
  {S.}~\bibnamefont {Araki}}, \bibinfo {author} {\bibfnamefont
  {A.}~\bibnamefont {Araya}}, \bibinfo {author} {\bibfnamefont
  {N.}~\bibnamefont {Aritomi}}, \bibinfo {author} {\bibfnamefont
  {Y.}~\bibnamefont {Aso}}, \bibinfo {author} {\bibfnamefont {S.}~\bibnamefont
  {Bae}}, \bibinfo {author} {\bibfnamefont {Y.}~\bibnamefont {Bae}}, \bibinfo
  {author} {\bibfnamefont {L.}~\bibnamefont {Baiotti}}, \bibinfo {author}
  {\bibfnamefont {R.}~\bibnamefont {Bajpai}}, \bibinfo {author} {\bibfnamefont
  {M.~A.}\ \bibnamefont {Barton}}, \bibinfo {author} {\bibfnamefont
  {K.}~\bibnamefont {Cannon}}, \bibinfo {author} {\bibfnamefont
  {E.}~\bibnamefont {Capocasa}}, \bibinfo {author} {\bibfnamefont
  {M.}~\bibnamefont {Chan}}, \bibinfo {author} {\bibfnamefont {C.}~\bibnamefont
  {Chen}}, \bibinfo {author} {\bibfnamefont {K.}~\bibnamefont {Chen}}, \bibinfo
  {author} {\bibfnamefont {Y.}~\bibnamefont {Chen}}, \bibinfo {author}
  {\bibfnamefont {H.}~\bibnamefont {Chu}}, \bibinfo {author} {\bibfnamefont
  {Y.~K.}\ \bibnamefont {Chu}}, \bibinfo {author} {\bibfnamefont
  {S.}~\bibnamefont {Eguchi}}, \bibinfo {author} {\bibfnamefont
  {Y.}~\bibnamefont {Enomoto}}, \bibinfo {author} {\bibfnamefont
  {R.}~\bibnamefont {Flaminio}}, \bibinfo {author} {\bibfnamefont
  {Y.}~\bibnamefont {Fujii}}, \bibinfo {author} {\bibfnamefont
  {M.}~\bibnamefont {Fukunaga}}, \bibinfo {author} {\bibfnamefont
  {M.}~\bibnamefont {Fukushima}}, \bibinfo {author} {\bibfnamefont
  {G.}~\bibnamefont {Ge}}, \bibinfo {author} {\bibfnamefont {A.}~\bibnamefont
  {Hagiwara}}, \bibinfo {author} {\bibfnamefont {S.}~\bibnamefont {Haino}},
  \bibinfo {author} {\bibfnamefont {K.}~\bibnamefont {Hasegawa}}, \bibinfo
  {author} {\bibfnamefont {H.}~\bibnamefont {Hayakawa}}, \bibinfo {author}
  {\bibfnamefont {K.}~\bibnamefont {Hayama}}, \bibinfo {author} {\bibfnamefont
  {Y.}~\bibnamefont {Himemoto}}, \bibinfo {author} {\bibfnamefont
  {Y.}~\bibnamefont {Hiranuma}}, \bibinfo {author} {\bibfnamefont
  {N.}~\bibnamefont {Hirata}}, \bibinfo {author} {\bibfnamefont
  {E.}~\bibnamefont {Hirose}}, \bibinfo {author} {\bibfnamefont
  {Z.}~\bibnamefont {Hong}}, \bibinfo {author} {\bibfnamefont {B.~H.}\
  \bibnamefont {Hsieh}}, \bibinfo {author} {\bibfnamefont {C.~Z.}\ \bibnamefont
  {Huang}}, \bibinfo {author} {\bibfnamefont {P.}~\bibnamefont {Huang}},
  \bibinfo {author} {\bibfnamefont {Y.}~\bibnamefont {Huang}}, \bibinfo
  {author} {\bibfnamefont {B.}~\bibnamefont {Ikenoue}}, \bibinfo {author}
  {\bibfnamefont {S.}~\bibnamefont {Imam}}, \bibinfo {author} {\bibfnamefont
  {K.}~\bibnamefont {Inayoshi}}, \bibinfo {author} {\bibfnamefont
  {Y.}~\bibnamefont {Inoue}}, \bibinfo {author} {\bibfnamefont
  {K.}~\bibnamefont {Ioka}}, \bibinfo {author} {\bibfnamefont {Y.}~\bibnamefont
  {Itoh}}, \bibinfo {author} {\bibfnamefont {K.}~\bibnamefont {Izumi}},
  \bibinfo {author} {\bibfnamefont {K.}~\bibnamefont {Jung}}, \bibinfo {author}
  {\bibfnamefont {P.}~\bibnamefont {Jung}}, \bibinfo {author} {\bibfnamefont
  {T.}~\bibnamefont {Kajita}}, \bibinfo {author} {\bibfnamefont
  {M.}~\bibnamefont {Kamiizumi}}, \bibinfo {author} {\bibfnamefont
  {N.}~\bibnamefont {Kanda}}, \bibinfo {author} {\bibfnamefont
  {G.}~\bibnamefont {Kang}}, \bibinfo {author} {\bibfnamefont {K.}~\bibnamefont
  {Kawaguchi}}, \bibinfo {author} {\bibfnamefont {N.}~\bibnamefont {Kawai}},
  \bibinfo {author} {\bibfnamefont {T.}~\bibnamefont {Kawasaki}}, \bibinfo
  {author} {\bibfnamefont {C.}~\bibnamefont {Kim}}, \bibinfo {author}
  {\bibfnamefont {J.~C.}\ \bibnamefont {Kim}}, \bibinfo {author} {\bibfnamefont
  {W.~S.}\ \bibnamefont {Kim}}, \bibinfo {author} {\bibfnamefont {Y.~M.}\
  \bibnamefont {Kim}}, \bibinfo {author} {\bibfnamefont {N.}~\bibnamefont
  {Kimura}}, \bibinfo {author} {\bibfnamefont {N.}~\bibnamefont {Kita}},
  \bibinfo {author} {\bibfnamefont {H.}~\bibnamefont {Kitazawa}}, \bibinfo
  {author} {\bibfnamefont {Y.}~\bibnamefont {Kojima}}, \bibinfo {author}
  {\bibfnamefont {K.}~\bibnamefont {Kokeyama}}, \bibinfo {author}
  {\bibfnamefont {K.}~\bibnamefont {Komori}}, \bibinfo {author} {\bibfnamefont
  {A.~K.~H.}\ \bibnamefont {Kong}}, \bibinfo {author} {\bibfnamefont
  {K.}~\bibnamefont {Kotake}}, \bibinfo {author} {\bibfnamefont
  {C.}~\bibnamefont {Kozakai}}, \bibinfo {author} {\bibfnamefont
  {R.}~\bibnamefont {Kozu}}, \bibinfo {author} {\bibfnamefont {R.}~\bibnamefont
  {Kumar}}, \bibinfo {author} {\bibfnamefont {J.}~\bibnamefont {Kume}},
  \bibinfo {author} {\bibfnamefont {C.}~\bibnamefont {Kuo}}, \bibinfo {author}
  {\bibfnamefont {H.~S.}\ \bibnamefont {Kuo}}, \bibinfo {author} {\bibfnamefont
  {S.}~\bibnamefont {Kuroyanagi}}, \bibinfo {author} {\bibfnamefont
  {K.}~\bibnamefont {Kusayanagi}}, \bibinfo {author} {\bibfnamefont
  {K.}~\bibnamefont {Kwak}}, \bibinfo {author} {\bibfnamefont {H.~K.}\
  \bibnamefont {Lee}}, \bibinfo {author} {\bibfnamefont {H.~W.}\ \bibnamefont
  {Lee}}, \bibinfo {author} {\bibfnamefont {R.}~\bibnamefont {Lee}}, \bibinfo
  {author} {\bibfnamefont {M.}~\bibnamefont {Leonardi}}, \bibinfo {author}
  {\bibfnamefont {L.~C.~C.}\ \bibnamefont {Lin}}, \bibinfo {author}
  {\bibfnamefont {C.~Y.}\ \bibnamefont {Lin}}, \bibinfo {author} {\bibfnamefont
  {F.~L.}\ \bibnamefont {Lin}}, \bibinfo {author} {\bibfnamefont {G.~C.}\
  \bibnamefont {Liu}}, \bibinfo {author} {\bibfnamefont {L.~W.}\ \bibnamefont
  {Luo}}, \bibinfo {author} {\bibfnamefont {M.}~\bibnamefont {Marchio}},
  \bibinfo {author} {\bibfnamefont {Y.}~\bibnamefont {Michimura}}, \bibinfo
  {author} {\bibfnamefont {N.}~\bibnamefont {Mio}}, \bibinfo {author}
  {\bibfnamefont {O.}~\bibnamefont {Miyakawa}}, \bibinfo {author}
  {\bibfnamefont {A.}~\bibnamefont {Miyamoto}}, \bibinfo {author}
  {\bibfnamefont {Y.}~\bibnamefont {Miyazaki}}, \bibinfo {author}
  {\bibfnamefont {K.}~\bibnamefont {Miyo}}, \bibinfo {author} {\bibfnamefont
  {S.}~\bibnamefont {Miyoki}}, \bibinfo {author} {\bibfnamefont
  {S.}~\bibnamefont {Morisaki}}, \bibinfo {author} {\bibfnamefont
  {Y.}~\bibnamefont {Moriwaki}}, \bibinfo {author} {\bibfnamefont
  {K.}~\bibnamefont {Nagano}}, \bibinfo {author} {\bibfnamefont
  {S.}~\bibnamefont {Nagano}}, \bibinfo {author} {\bibfnamefont
  {K.}~\bibnamefont {Nakamura}}, \bibinfo {author} {\bibfnamefont
  {H.}~\bibnamefont {Nakano}}, \bibinfo {author} {\bibfnamefont
  {M.}~\bibnamefont {Nakano}}, \bibinfo {author} {\bibfnamefont
  {R.}~\bibnamefont {Nakashima}}, \bibinfo {author} {\bibfnamefont
  {T.}~\bibnamefont {Narikawa}}, \bibinfo {author} {\bibfnamefont
  {R.}~\bibnamefont {Negishi}}, \bibinfo {author} {\bibfnamefont {W.~T.}\
  \bibnamefont {Ni}}, \bibinfo {author} {\bibfnamefont {A.}~\bibnamefont
  {Nishizawa}}, \bibinfo {author} {\bibfnamefont {Y.}~\bibnamefont {Obuchi}},
  \bibinfo {author} {\bibfnamefont {W.}~\bibnamefont {Ogaki}}, \bibinfo
  {author} {\bibfnamefont {J.~J.}\ \bibnamefont {Oh}}, \bibinfo {author}
  {\bibfnamefont {S.~H.}\ \bibnamefont {Oh}}, \bibinfo {author} {\bibfnamefont
  {M.}~\bibnamefont {Ohashi}}, \bibinfo {author} {\bibfnamefont
  {N.}~\bibnamefont {Ohishi}}, \bibinfo {author} {\bibfnamefont
  {M.}~\bibnamefont {Ohkawa}}, \bibinfo {author} {\bibfnamefont
  {K.}~\bibnamefont {Okutomi}}, \bibinfo {author} {\bibfnamefont
  {K.}~\bibnamefont {Oohara}}, \bibinfo {author} {\bibfnamefont {C.~P.}\
  \bibnamefont {Ooi}}, \bibinfo {author} {\bibfnamefont {S.}~\bibnamefont
  {Oshino}}, \bibinfo {author} {\bibfnamefont {K.}~\bibnamefont {Pan}},
  \bibinfo {author} {\bibfnamefont {H.}~\bibnamefont {Pang}}, \bibinfo {author}
  {\bibfnamefont {J.}~\bibnamefont {Park}}, \bibinfo {author} {\bibfnamefont
  {F.~E.~P.}\ \bibnamefont {Arellano}}, \bibinfo {author} {\bibfnamefont
  {I.}~\bibnamefont {Pinto}}, \bibinfo {author} {\bibfnamefont
  {N.}~\bibnamefont {Sago}}, \bibinfo {author} {\bibfnamefont {S.}~\bibnamefont
  {Saito}}, \bibinfo {author} {\bibfnamefont {Y.}~\bibnamefont {Saito}},
  \bibinfo {author} {\bibfnamefont {K.}~\bibnamefont {Sakai}}, \bibinfo
  {author} {\bibfnamefont {Y.}~\bibnamefont {Sakai}}, \bibinfo {author}
  {\bibfnamefont {Y.}~\bibnamefont {Sakuno}}, \bibinfo {author} {\bibfnamefont
  {S.}~\bibnamefont {Sato}}, \bibinfo {author} {\bibfnamefont {T.}~\bibnamefont
  {Sato}}, \bibinfo {author} {\bibfnamefont {T.}~\bibnamefont {Sawada}},
  \bibinfo {author} {\bibfnamefont {T.}~\bibnamefont {Sekiguchi}}, \bibinfo
  {author} {\bibfnamefont {Y.}~\bibnamefont {Sekiguchi}}, \bibinfo {author}
  {\bibfnamefont {S.}~\bibnamefont {Shibagaki}}, \bibinfo {author}
  {\bibfnamefont {R.}~\bibnamefont {Shimizu}}, \bibinfo {author} {\bibfnamefont
  {T.}~\bibnamefont {Shimoda}}, \bibinfo {author} {\bibfnamefont
  {K.}~\bibnamefont {Shimode}}, \bibinfo {author} {\bibfnamefont
  {H.}~\bibnamefont {Shinkai}}, \bibinfo {author} {\bibfnamefont
  {T.}~\bibnamefont {Shishido}}, \bibinfo {author} {\bibfnamefont
  {A.}~\bibnamefont {Shoda}}, \bibinfo {author} {\bibfnamefont
  {K.}~\bibnamefont {Somiya}}, \bibinfo {author} {\bibfnamefont {E.~J.}\
  \bibnamefont {Son}}, \bibinfo {author} {\bibfnamefont {H.}~\bibnamefont
  {Sotani}}, \bibinfo {author} {\bibfnamefont {R.}~\bibnamefont {Sugimoto}},
  \bibinfo {author} {\bibfnamefont {T.}~\bibnamefont {Suzuki}}, \bibinfo
  {author} {\bibfnamefont {T.}~\bibnamefont {Suzuki}}, \bibinfo {author}
  {\bibfnamefont {H.}~\bibnamefont {Tagoshi}}, \bibinfo {author} {\bibfnamefont
  {H.}~\bibnamefont {Takahashi}}, \bibinfo {author} {\bibfnamefont
  {R.}~\bibnamefont {Takahashi}}, \bibinfo {author} {\bibfnamefont
  {A.}~\bibnamefont {Takamori}}, \bibinfo {author} {\bibfnamefont
  {S.}~\bibnamefont {Takano}}, \bibinfo {author} {\bibfnamefont
  {H.}~\bibnamefont {Takeda}}, \bibinfo {author} {\bibfnamefont
  {M.}~\bibnamefont {Takeda}}, \bibinfo {author} {\bibfnamefont
  {H.}~\bibnamefont {Tanaka}}, \bibinfo {author} {\bibfnamefont
  {K.}~\bibnamefont {Tanaka}}, \bibinfo {author} {\bibfnamefont
  {K.}~\bibnamefont {Tanaka}}, \bibinfo {author} {\bibfnamefont
  {T.}~\bibnamefont {Tanaka}}, \bibinfo {author} {\bibfnamefont
  {T.}~\bibnamefont {Tanaka}}, \bibinfo {author} {\bibfnamefont
  {S.}~\bibnamefont {Tanioka}}, \bibinfo {author} {\bibfnamefont {E.~N.}\
  \bibnamefont {Tapia San~Martin}}, \bibinfo {author} {\bibfnamefont
  {S.}~\bibnamefont {Telada}}, \bibinfo {author} {\bibfnamefont
  {T.}~\bibnamefont {Tomaru}}, \bibinfo {author} {\bibfnamefont
  {Y.}~\bibnamefont {Tomigami}}, \bibinfo {author} {\bibfnamefont
  {T.}~\bibnamefont {Tomura}}, \bibinfo {author} {\bibfnamefont
  {F.}~\bibnamefont {Travasso}}, \bibinfo {author} {\bibfnamefont
  {L.}~\bibnamefont {Trozzo}}, \bibinfo {author} {\bibfnamefont
  {T.}~\bibnamefont {Tsang}}, \bibinfo {author} {\bibfnamefont
  {K.}~\bibnamefont {Tsubono}}, \bibinfo {author} {\bibfnamefont
  {S.}~\bibnamefont {Tsuchida}}, \bibinfo {author} {\bibfnamefont
  {T.}~\bibnamefont {Tsuzuki}}, \bibinfo {author} {\bibfnamefont
  {D.}~\bibnamefont {Tuyenbayev}}, \bibinfo {author} {\bibfnamefont
  {N.}~\bibnamefont {Uchikata}}, \bibinfo {author} {\bibfnamefont
  {T.}~\bibnamefont {Uchiyama}}, \bibinfo {author} {\bibfnamefont
  {A.}~\bibnamefont {Ueda}}, \bibinfo {author} {\bibfnamefont {T.}~\bibnamefont
  {Uehara}}, \bibinfo {author} {\bibfnamefont {K.}~\bibnamefont {Ueno}},
  \bibinfo {author} {\bibfnamefont {G.}~\bibnamefont {Ueshima}}, \bibinfo
  {author} {\bibfnamefont {F.}~\bibnamefont {Uraguchi}}, \bibinfo {author}
  {\bibfnamefont {T.}~\bibnamefont {Ushiba}}, \bibinfo {author} {\bibfnamefont
  {M.~H. P.~M.}\ \bibnamefont {van Putten}}, \bibinfo {author} {\bibfnamefont
  {H.}~\bibnamefont {Vocca}}, \bibinfo {author} {\bibfnamefont
  {J.}~\bibnamefont {Wang}}, \bibinfo {author} {\bibfnamefont {C.}~\bibnamefont
  {Wu}}, \bibinfo {author} {\bibfnamefont {H.}~\bibnamefont {Wu}}, \bibinfo
  {author} {\bibfnamefont {S.}~\bibnamefont {Wu}}, \bibinfo {author}
  {\bibfnamefont {W.-R.}\ \bibnamefont {Xu}}, \bibinfo {author} {\bibfnamefont
  {T.}~\bibnamefont {Yamada}}, \bibinfo {author} {\bibfnamefont
  {K.}~\bibnamefont {Yamamoto}}, \bibinfo {author} {\bibfnamefont
  {K.}~\bibnamefont {Yamamoto}}, \bibinfo {author} {\bibfnamefont
  {T.}~\bibnamefont {Yamamoto}}, \bibinfo {author} {\bibfnamefont
  {K.}~\bibnamefont {Yokogawa}}, \bibinfo {author} {\bibfnamefont
  {J.}~\bibnamefont {Yokoyama}}, \bibinfo {author} {\bibfnamefont
  {T.}~\bibnamefont {Yokozawa}}, \bibinfo {author} {\bibfnamefont
  {T.}~\bibnamefont {Yoshioka}}, \bibinfo {author} {\bibfnamefont
  {H.}~\bibnamefont {Yuzurihara}}, \bibinfo {author} {\bibfnamefont
  {S.}~\bibnamefont {Zeidler}}, \bibinfo {author} {\bibfnamefont
  {Y.}~\bibnamefont {Zhao}},\ and\ \bibinfo {author} {\bibfnamefont {Z.~H.}\
  \bibnamefont {Zhu}},\ }\bibfield  {journal} {\bibinfo  {journal} {Progress of
  Theoretical and Experimental Physics}\ }\textbf {\bibinfo {volume} {2021}},\
  \href {https://doi.org/10.1093/ptep/ptaa125} {10.1093/ptep/ptaa125} (\bibinfo
  {year} {2020}),\ \bibinfo {note} {05A101},\ \Eprint
  {https://arxiv.org/abs/https://academic.oup.com/ptep/article-pdf/2021/5/05A101/37974994/ptaa125.pdf}
  {https://academic.oup.com/ptep/article-pdf/2021/5/05A101/37974994/ptaa125.pdf}
  \BibitemShut {NoStop}%
\bibitem [{\citenamefont {Reid}\ and\ \citenamefont
  {Martin}(2016)}]{coatings6040061}%
  \BibitemOpen
  \bibfield  {author} {\bibinfo {author} {\bibfnamefont {S.}~\bibnamefont
  {Reid}}\ and\ \bibinfo {author} {\bibfnamefont {I.~W.}\ \bibnamefont
  {Martin}},\ }\bibfield  {journal} {\bibinfo  {journal} {Coatings}\ }\textbf
  {\bibinfo {volume} {6}},\ \href {https://doi.org/10.3390/coatings6040061}
  {10.3390/coatings6040061} (\bibinfo {year} {2016})\BibitemShut {NoStop}%
\bibitem [{\citenamefont {Sidqi}\ \emph {et~al.}(2019)\citenamefont {Sidqi},
  \citenamefont {Clark}, \citenamefont {Buller}, \citenamefont {Thalluri},
  \citenamefont {Mitrofanov},\ and\ \citenamefont {Noblet}}]{Sidqi:19}%
  \BibitemOpen
  \bibfield  {author} {\bibinfo {author} {\bibfnamefont {N.}~\bibnamefont
  {Sidqi}}, \bibinfo {author} {\bibfnamefont {C.}~\bibnamefont {Clark}},
  \bibinfo {author} {\bibfnamefont {G.~S.}\ \bibnamefont {Buller}}, \bibinfo
  {author} {\bibfnamefont {G.~K. V.~V.}\ \bibnamefont {Thalluri}}, \bibinfo
  {author} {\bibfnamefont {J.}~\bibnamefont {Mitrofanov}},\ and\ \bibinfo
  {author} {\bibfnamefont {Y.}~\bibnamefont {Noblet}},\ }\href
  {https://doi.org/10.1364/OME.9.003452} {\bibfield  {journal} {\bibinfo
  {journal} {Opt. Mater. Express}\ }\textbf {\bibinfo {volume} {9}},\ \bibinfo
  {pages} {3452} (\bibinfo {year} {2019})}\BibitemShut {NoStop}%
\bibitem [{\citenamefont {MILOTTI}\ \emph {et~al.}(2012)\citenamefont
  {MILOTTI}, \citenamefont {DELLA~VALLE}, \citenamefont {ZAVATTINI},
  \citenamefont {MESSINEO}, \citenamefont {GASTALDI}, \citenamefont {PENGO},
  \citenamefont {RUOSO}, \citenamefont {BABUSCI}, \citenamefont {CURCEANU},
  \citenamefont {ILIESCU},\ and\ \citenamefont
  {MILARDI}}]{doi:10.1142S021974991241002X}%
  \BibitemOpen
  \bibfield  {author} {\bibinfo {author} {\bibfnamefont {E.}~\bibnamefont
  {MILOTTI}}, \bibinfo {author} {\bibfnamefont {F.}~\bibnamefont
  {DELLA~VALLE}}, \bibinfo {author} {\bibfnamefont {G.}~\bibnamefont
  {ZAVATTINI}}, \bibinfo {author} {\bibfnamefont {G.}~\bibnamefont {MESSINEO}},
  \bibinfo {author} {\bibfnamefont {U.}~\bibnamefont {GASTALDI}}, \bibinfo
  {author} {\bibfnamefont {R.}~\bibnamefont {PENGO}}, \bibinfo {author}
  {\bibfnamefont {G.}~\bibnamefont {RUOSO}}, \bibinfo {author} {\bibfnamefont
  {D.}~\bibnamefont {BABUSCI}}, \bibinfo {author} {\bibfnamefont
  {C.}~\bibnamefont {CURCEANU}}, \bibinfo {author} {\bibfnamefont
  {M.}~\bibnamefont {ILIESCU}},\ and\ \bibinfo {author} {\bibfnamefont
  {C.}~\bibnamefont {MILARDI}},\ }\href
  {https://doi.org/10.1142/S021974991241002X} {\bibfield  {journal} {\bibinfo
  {journal} {International Journal of Quantum Information}\ }\textbf {\bibinfo
  {volume} {10}},\ \bibinfo {pages} {1241002} (\bibinfo {year} {2012})},\
  \Eprint {https://arxiv.org/abs/https://doi.org/10.1142/S021974991241002X}
  {https://doi.org/10.1142/S021974991241002X} \BibitemShut {NoStop}%
\bibitem [{\citenamefont {Valle}\ \emph {et~al.}(2013)\citenamefont {Valle},
  \citenamefont {Gastaldi}, \citenamefont {Messineo}, \citenamefont {Milotti},
  \citenamefont {Pengo}, \citenamefont {Piemontese}, \citenamefont {Ruoso},\
  and\ \citenamefont {Zavattini}}]{Della_Valle_2013}%
  \BibitemOpen
  \bibfield  {author} {\bibinfo {author} {\bibfnamefont {F.~D.}\ \bibnamefont
  {Valle}}, \bibinfo {author} {\bibfnamefont {U.}~\bibnamefont {Gastaldi}},
  \bibinfo {author} {\bibfnamefont {G.}~\bibnamefont {Messineo}}, \bibinfo
  {author} {\bibfnamefont {E.}~\bibnamefont {Milotti}}, \bibinfo {author}
  {\bibfnamefont {R.}~\bibnamefont {Pengo}}, \bibinfo {author} {\bibfnamefont
  {L.}~\bibnamefont {Piemontese}}, \bibinfo {author} {\bibfnamefont
  {G.}~\bibnamefont {Ruoso}},\ and\ \bibinfo {author} {\bibfnamefont
  {G.}~\bibnamefont {Zavattini}},\ }\href
  {https://doi.org/10.1088/1367-2630/15/5/053026} {\bibfield  {journal}
  {\bibinfo  {journal} {New Journal of Physics}\ }\textbf {\bibinfo {volume}
  {15}},\ \bibinfo {pages} {053026} (\bibinfo {year} {2013})}\BibitemShut
  {NoStop}%
\bibitem [{\citenamefont {Shukla}\ \emph {et~al.}(2004)\citenamefont {Shukla},
  \citenamefont {Marklund}, \citenamefont {Tskhakaya},\ and\ \citenamefont
  {Eliasson}}]{doi:10.1063/1.1759628}%
  \BibitemOpen
  \bibfield  {author} {\bibinfo {author} {\bibfnamefont {P.~K.}\ \bibnamefont
  {Shukla}}, \bibinfo {author} {\bibfnamefont {M.}~\bibnamefont {Marklund}},
  \bibinfo {author} {\bibfnamefont {D.~D.}\ \bibnamefont {Tskhakaya}},\ and\
  \bibinfo {author} {\bibfnamefont {B.}~\bibnamefont {Eliasson}},\ }\href
  {https://doi.org/10.1063/1.1759628} {\bibfield  {journal} {\bibinfo
  {journal} {Physics of Plasmas}\ }\textbf {\bibinfo {volume} {11}},\ \bibinfo
  {pages} {3767} (\bibinfo {year} {2004})},\ \Eprint
  {https://arxiv.org/abs/https://doi.org/10.1063/1.1759628}
  {https://doi.org/10.1063/1.1759628} \BibitemShut {NoStop}%
\end{thebibliography}

%

\end{document}